\documentclass[twocolumn]{aastex631}

\usepackage{amsmath}
\usepackage{comment}

\newcommand{\Msun}{{\rm M_{\odot}}}

\newcommand{\Zsun}{{\rm Z_{\odot}}}
\newcommand{\mpc}{\, {\rm Mpc}}
\newcommand{\kpc}{\, {\rm kpc}}
\newcommand{\pc}{\, {\rm pc}}
\newcommand{\kmps}{\, {\rm km \, s^{-1}}}

\shorttitle{CO(3-2) Emission in M83's XUV Disk}
\shortauthors{Koda et al.}

\begin{document}

\title{First Detection of the Molecular Cloud Population in the Extended Ultraviolet (XUV) Disk of M83 }

\correspondingauthor{Jin Koda}
\email{jin.koda@stonybrook.edu}

\author{Jin Koda}
\affil{Department of Physics and Astronomy, Stony Brook University, Stony Brook, NY 11794-3800}

\author{Linda Watson}
\affil{European Southern Observatory; Joint ALMA Observatory}

\author{Fran\c{c}oise Combes}
\affil{Observatoire de Paris, LERMA, Coll\`ege de France,  PSL Univ., CNRS, Sorbonne Univ., Paris, France}

\author[0000-0002-5307-5941]{Monica Rubio}
\affil{Departamento de Astronomia, Universidad de Chile, Casilla 36-D, Santiago, Chile}

\author{Samuel Boissier}
\affil{Aix Marseille Univ., CNRS, CNES, Laboratoire d'Astrophysique de Marseille, Marseille, France}

\author{Masafumi Yagi}
\affil{National Astronomical Observatory of Japan, Mitaka, Tokyo, 181-8588, Japan}

\author{David Thilker}
\affil{Department of Physics and Astronomy, The Johns Hopkins University, Baltimore, MD 21218, USA}

\author{Amanda M Lee}
\affil{Department of Physics and Astronomy, Stony Brook University, Stony Brook, NY 11794-3800}

\author{Yutaka Komiyama}
\affil{Dept. of Advanced Sciences, Faculty of Science and Engineering, Hosei University,
3-7-2 Kajino-cho, Koganei-shi, Tokyo 184-8584, Japan}
\affil{National Astronomical Observatory of Japan, Mitaka, Tokyo, 181-8588, Japan}
\affil{Graduate University for Advanced Studies (SOKENDAI), Mitaka, Tokyo 181-8588, Japan}

\author[0000-0003-3932-0952]{Kana Morokuma-Matsui}
\affiliation{Institute of Astronomy, Graduate School of Science, The
University of Tokyo, 2-21-1 Osawa, Mitaka, Tokyo 181-0015, Japan}

\author{Celia Verdugo}
\affil{Joint ALMA Observatory (JAO), Alonso de C\'ordova 3107, Vitacura, Santiago, Chile}

\begin{abstract}
We report a CO($J$=3-2) detection of 23 molecular clouds
in the extended ultraviolet (XUV) disk of the spiral galaxy M83
with the Atacama Large Millimeter/submillimeter Array (ALMA).
The observed $1 \kpc^2$ region is at about 1.24 times the optical radius ($R_{25}$) of the disk,
where CO($J$=2-1) was previously not detected.
The detection and non-detection, as well as the level of star formation (SF) activity in the region,
can be explained consistently if the clouds have
the mass distribution common among Galactic clouds, such as Orion A
-- with star-forming dense clumps embedded in thick layers of bulk molecular gas,
but in a low-metallicity regime where their outer layers are CO-deficient and CO-dark.
The cloud and clump masses, estimated from CO(3-2),
range from $8.2\times 10^2$ to $2.3\times 10^4\Msun$
and from $2.7\times 10^2$ to $7.5\times 10^3\Msun$, respectively.
The most massive clouds appear similar to Orion A in star formation activity
as well as in mass, as expected if the cloud mass structure is common.
The overall low SF activity in the XUV disk could be
due to the relative shortage of gas in the molecular phase.
The clouds are distributed like chains up to 600~pc (or longer) in length, 
suggesting that the trigger of cloud formation is on large scales.
The common cloud mass structure also justifies the use of high-$J$ CO transitions
to trace the total gas mass of clouds, or galaxies, even in the high-$z$ universe.
This study is the first demonstration that CO(3-2) is an efficient tracer
of molecular clouds even in low-metallicity environments.
\end{abstract}


\section{Introduction} \label{sec:intro}

\begin{figure*}[b]
\epsscale{1.0}
\plotone{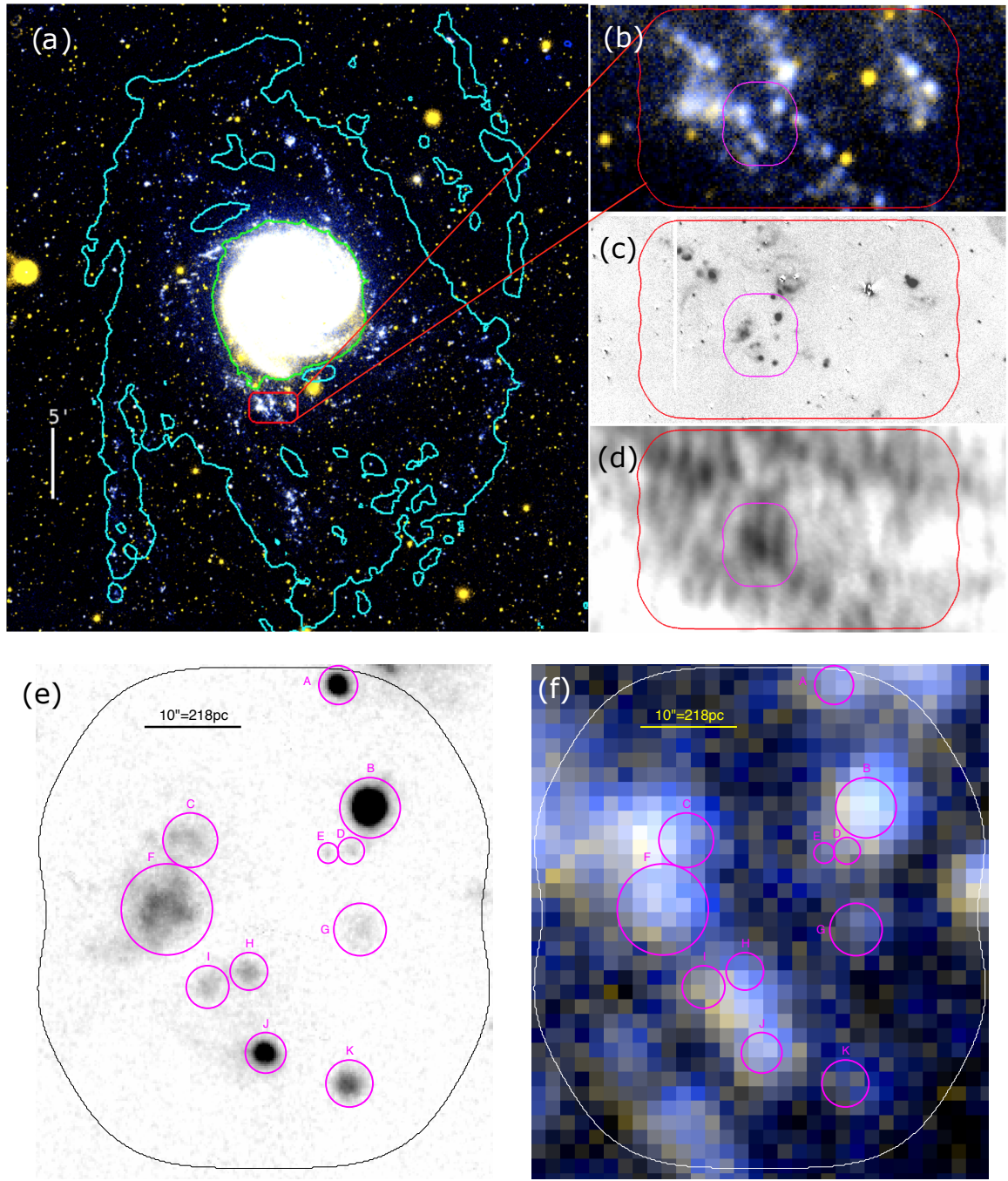}
\caption{
(a) GALEX FUV \& NUV-band color composite image. The inner (green) and outer (cyan) contours are the edge of the optical disk and the extent of HI gas at an HI surface density of $1.5\times 10^{20}\,\rm cm^{-2}$ \citep[see ][]{Koda:2012ab}. The red box is the $1.5\arcmin \times 3\arcmin$ ($2.0 \times 3.9\kpc^2$) field from the ALMA CO($J$=2-1) observations \citep{Bicalho:2019aa}.
The three panels on the right are zoom-in to the CO(2-1) field:
(b) the GALEX color composite, as in Panel a,
(c) the continuum-subtracted H$\alpha$ image from the Subaru telescope, and
(d) the HI 21cm emission image from \cite{Walter:2008mw}.
The area of our CO(3-2) observations is indicated by the inner magenta box of $\sim 0.75\arcmin \times 0.85\arcmin$ ($\sim 0.98 \times 1.11 \kpc^2$).
The bottom two panels are zoom-in to the CO(3-2) field:
(e) The H$\alpha$ image with visually-identified HII regions (circles).
(f) The same as (e), but on the GALEX color composite.
Panel e,f, Figures \ref{fig:maps}, \ref{fig:chmap32cont}, and \ref{fig:UVHaCO} show the same region, and their coordinates are explicitly written in Figures \ref{fig:maps} and \ref{fig:chmap32cont}.
\label{fig:fov}}
\end{figure*}

The Galaxy Evolution Explorer (GALEX) satellite found 
massive star formation (SF) in the far outskirts
of galactic disks \citep{Gil-de-Paz:2005aa, Thilker:2005ff}.
Bright ultraviolet (UV) sources are distributed beyond
the optical radius $R_{25}$.
They reveal abundant and recent ($<100$~Myr) SF
\citep[][see Figure \ref{fig:fov} for the example of M83]{Gil-de-Paz:2005aa, Thilker:2005ff, Lemonias:2011aa}.
These extended UV disks, dubbed XUV disks, are fairly common
among local disk galaxies \citep{Thilker:2007dp}.
They offer an opportunity to study SF in extreme conditions,
in particular, at a low average gas density and molecular fraction.
XUV disks often exhibit lower H$\alpha$-to-far UV (FUV) flux ratios
compared to the optical disk, which likely demonstrates the importance of stochastic sampling of the initial mass function (IMF) and/or bursty SF histories in these regions
\citep[e.g., ][]{Alberts:2011aa, Koda:2012ab, Watson:2016aa}.

Molecular clouds host virtually all SF within the optical disks of local galaxies.
It is crucial to study whether the same is true in XUV disks.
The Milky Way's outer disk hosts small molecular clouds \citep[$10^{2-4}\Msun$; ][]{Sun:2015wr, Sun:2017vv}, but unlike XUV disks there is not much SF above late B-type stars \citep{Izumi:2017wv}.
Numerous efforts have been made to detect CO emission in XUV disks, however,
they have rarely succeeded
\citep[][for review including unpublished efforts that resulted in non-detections]{Watson:2017aa}.
Only four galaxies with XUV disks permitted CO detection at a few positions in their outskirts:
NGC 4414 \citep{Braine:2004aa},
NGC 6946 \citep{Braine:2007aa},
M33 \citep{Braine:2010aa}, and
M63 \citep{Dessauges-Zavadsky:2014fk}.
Most detections could not reveal
individual clouds in the vicinity of UV emission
due to low sensitivity and resolution 
($\gtrsim 10^{5-6}\Msun$ and $\gtrsim 300$-$500\pc$).
Only M33, the closest to the Milky Way, allowed detection
down to $\gtrsim 4\times 10^4\Msun$ in two regions beyond the $R_{25}$ radius,
and only one of them is identified as a single cloud \citep{Braine:2012aa}.
The rarity of the CO detection is at odds with the abundance of the UV sources
across the XUV disks.

M83 is at a distance of $D=4.5\mpc$ \citep{Thim:2003aa} and is one of the nearest XUV disks.
Previous ALMA observations of CO(2-1) in the XUV disk of M83 resulted in non-detection \citep{Bicalho:2019aa}
even at a high sensitivity of $2.2\times 10^4\Msun$
\citep[$3\sigma$; calculated with the low CO-to-H$_2$ conversion
factor in the Large Magellanic Cloud (LMC) from ][]{Fukui:2008tl}.
The observed field is a relatively large area of $\sim 1.5\arcmin \times 3 \arcmin$ (2 $\times$ 4 kpc$^2$)
at a galactic radius of $r_{\rm gal}\sim 1.24 R_{25}$
with $R_{25}=6.44\arcmin$ ($=D_{25}/2$) from the optical center at 
(RA, DEC)$_{\rm J2000}$ = (13:37:00.4, -29:52:04.1) \citep{de-Vaucouleurs:1991lr}.
We revisit a smaller area of $\sim 0.75\arcmin \times 0.85\arcmin$
within the CO(2-1) field at a higher mass sensitivity
by a factor of about 10 in the CO(3-2) line emission,
as well as in the Band~7 dust continuum emission.

M83 presents a prototype XUV disk \citep{Thilker:2005ff}.
Deep H$\alpha$ imaging with the Subaru telescope revealed
HII regions over the XUV disk,
including some in the region of this study \citep{Koda:2012ab}.
The metallicity is about constant across the XUV disk and is as low as 12 + log(O/H) $\sim$ 8.2 \citep[``preferred abundances" by ][ from their analyses with multiple metallicity indicators]{Bresolin:2009ce}
\footnote{The scatters in measured metallicities are large among adopted metallicity indicators and among individual HII regions.
Within our ALMA coverage, \citet{Bresolin:2009ce} could apply the direct method,
the most reliable indicator, to only one HII region (\#39 in their catalog)
and obtain 12 + log(O/H) $\sim$ 8.05$\pm$0.07 ($\sim 0.3\Zsun$).
\citet{Bresolin:2009ce} also obtained 8.24, 8.19, and 8.45 with the [NII]/[OII] indicator, and 8.34, 8.37, and 8.43 with the N2 indicator for HII regions \#39, 40, and 41, respectively.
They analyzed much larger areas than our ALMA coverage,
and from totality, suggested a constant metallicity of
12 + log(O/H) $\sim$ 8.2 for the full XUV disk.}.
This is $\sim 0.4\Zsun$ by adopting the solar abundance of 12 + log(O/H)$_{\odot}$=8.66 \citep{Asplund:2005uv},
which is similar to that of the LMC \citep[$\sim 0.3$-$0.5\Zsun$, ][]{Westerlund:1997va}.
With Spitzer infrared images,
\citet{Dong:2008wb} concluded that
the SF has been ongoing for at least 1~Gyr.
The SF efficiency derived with
atomic hydrogen (HI) data indicates a gas consumption timescale
much longer than the Hubble time \citep{Bigiel:2010yq}.
This extremely low SF efficiency could be due to
physical processes either on large scales (e.g., low HI-to-H$_2$ phase transition),
or on small scales (e.g., low SF rate within molecular clouds).
Observations and detection of molecular gas would improve
our understanding of SF in this extreme environment.

\section{Observations and Data}\label{sec:data}

\subsection{ALMA Observations}\label{sec:alma}
We observed a region of $\sim 0.75\arcmin \times 0.85\arcmin$
($\sim 0.98 \times 1.11 \kpc^2$; approximately $\sim 1\kpc^2$)
in the outskirts of M83 with ALMA's 12m array in CO(3-2) and
Band~7 dust continuum emission.
Figure \ref{fig:fov} shows the target region:
the left panel displays the global context of the galaxy,
and the right panels show the observed region with magenta boxes (smaller boxes).
The outer (cyan) contour in the left panel shows the extent of the HI disk,
which also encloses the XUV disk and numerous UV blobs (the color image).
The red boxes in all panels show the region of
the previous CO(2-1) and Band~6 continuum emission data \citep{Bicalho:2019aa}.
Our region is selected around the maximum HI 21~cm intensity position
within the CO(2-1) observation.
It includes bright UV peaks located
at the galactocentric distance of $\sim 1.24 R_{25}$ ($\sim 8.0\arcmin$, $\sim 10.4$~kpc),
and is outside the traditional optical disk (green contour).

The CO(3-2) line ($\nu_{\rm CO(3-2)}=345.79599$~GHz)
and continuum emission in Band~7 were observed
with ALMA (project \# 2017.1.00065.S).
It mosaicked the target region
with 17 pointing positions around the map center at (RA, DEC)$_{J2000}$ = (13:37:05.8, -29:59:57.4).
One spectral window (SPW) was configured for the line emission with
a bandwidth of 1.875~GHz ($1626\kmps$) with 1920 channels of 976.6~kHz width ($0.8466\kmps$).
The other three SPWs were configured to cover different sky frequencies
for the continuum emission with a bandwidth of 2~GHz with 128 channels per each SPW.
The central frequency of the continuum emission is 349.498~GHz (hereafter, 349~GHz continuum emission).
The full mosaic observations were repeated six times,
resulting in six execution blocks.
The final $uv$-coverage at the CO(3-2) frequency extends over the angular scales of 0.57-13.7$\arcsec$ ($\sim$12-299~pc).

The CO(2-1) line ($\nu_{\rm CO(2-1)}=230.53800$~GHz) and
continuum emission in Band~6 were taken in project \# 2013.1.00861.S and
were re-reduced for consistency within this study.
These data have been studied by \citet{Bicalho:2019aa},
and the details of the observations are found there.
The rectangle $1.5\arcmin \times 3\arcmin$ region
includes the area of the CO(3-2) and Band~7 continuum data.
The bandwidth and channel width are 1.875~GHz ($2,438\kmps$)
and 1.953~MHz ($2.540\kmps$) for the CO(2-1) emission.
The central frequency for the continuum emission is 224.516~GHz (hereafter, 225~GHz continuum emission).
The covered angular scales are 0.49-10.7$\arcsec$ ($\sim$11-233~pc) at the CO(2-1) frequency.

These data are calibrated using the data reduction scripts provided by the ALMA observatory
using the Common Astronomy Software Application \citep[CASA: ][]{McMullin:2007aa}.
The amplitude and phase of bandpass and gain calibrators are confirmed to be
flat over time and frequency after the calibrations.

For the CO(3-2) and CO(2-1) lines we generate data cubes with the \textsc{tclean} task in a standard way.
We use a cell size of $0.15\arcsec$ and channel width of $2.54\kmps$ (i.e., the channel width of
the CO(2-1) data) over a velocity range of $\sim$400 to 700$\kmps$.
While we use the $2.54\kmps$ cube throughout this study unless otherwise specified,
we also make a separate CO(3-2) data cube with a channel width of $1.00\kmps$.
This cube is used solely for a measurement of velocity dispersions of molecular clouds.
The 349 and 225~GHz continuum data are imaged with the multi frequency synthesis (MFS) mode of \textsc{tclean}.
The CO(2-1) and 225~GHz continuum data are regridded to
the image format of the CO(3-2) and 349~GHz continuum data for comparisons.
The parameters of the data are listed in Table \ref{tab:data},
including the total band width for continuum, 
the channel width for lines,
the cell/pixel size, beam size (its major and minor axis diameters, $b_{\rm maj}$, $b_{\rm min}$, and position angle PA), and root-mean-square (RMS) noise.

\floattable
\begin{deluxetable}{cccccccccc}
\tablecaption{Parameters of Reduced Data\label{tab:data}}
\tablehead{
\colhead{Data} & \colhead{Band Width}  & \colhead{Chan Width} & \colhead{Cell Size} & \colhead{Beam Size} & \nocolhead{}& \colhead{Covered Scale}  & \nocolhead{}& \multicolumn{2}{c}{RMS ($1\sigma$)}\\
\cline{5-5} \cline{7-7} \cline{9-10}
\nocolhead{} & \nocolhead{}  & \nocolhead{} & \nocolhead{}  & \colhead{$b_{\rm maj}$, $b_{\rm min}$, PA} & \nocolhead{}& \colhead{min, max}  & \nocolhead{}& \multicolumn{2}{c}{}\\
\nocolhead{} & \colhead{(GHz)}  & \colhead{($\kmps$)} & \colhead{($\arcsec$)} & \colhead{($\arcsec \times \arcsec$, $\arcdeg$)} & \nocolhead{}& \colhead{($\arcsec$, $\arcsec$)} & \nocolhead{}& \colhead{(mJy/beam)} & \colhead{(mK)}
}
\startdata
CO 3-2        &       & 2.54 & 0.15 & 0.96, 0.82, -84.7 &&0.57, 13.7&& 0.96  & 12.5 \\
              &       & 1.00 & 0.15 & 0.96, 0.82, -84.7 &&0.57, 13.7&& 1.31  & 17.0 \\
349~GHz Cont. & 5.625 &      & 0.15 & 0.95, 0.81, -85.0 &&0.56, 13.8&& 0.030 & 0.39\\
CO 2-1        &       & 2.54 & 0.15 & 0.76, 0.55, -88.6 &&0.49, 10.7&& 10.6  & 58\\
225~GHz Cont. & 5.625 &      & 0.15 & 0.79, 0.60, -81.9 &&0.47, 11.1&& 0.22  & 12
\enddata
\end{deluxetable}

\begin{figure*}[b]
\epsscale{1.08}
\plotone{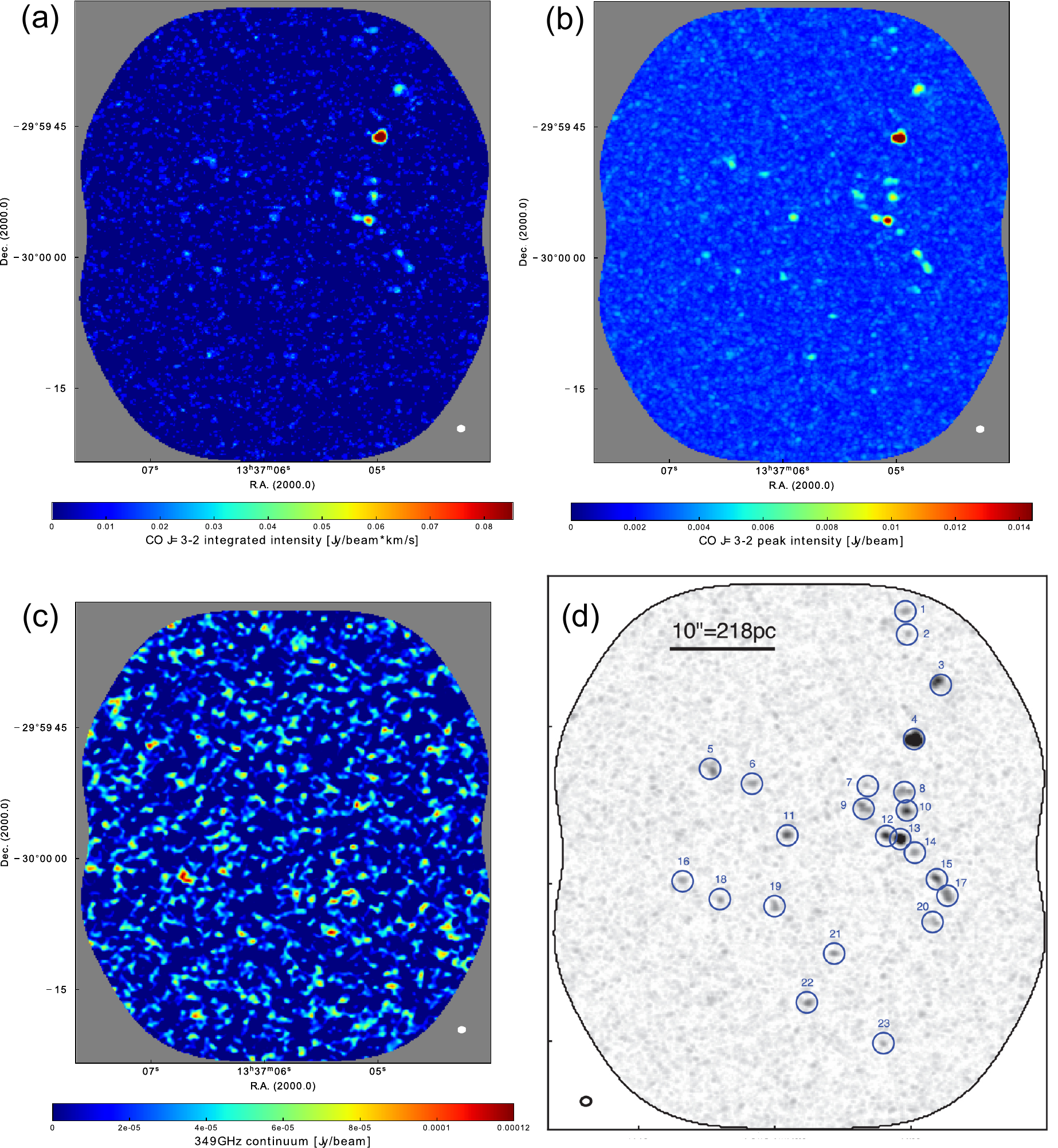}
\caption{
(a) CO(3-2) integrated intensity map.
(b) CO(3-2) peak intensity map.
(c) Band~7 (349~GHz) continuum map.
(d) Molecular clouds on the CO(3-2) peak intensity map.
Cloud IDs are indicated and correspond to the ones in Table \ref{tab:clouds}.
Each circle has a diameter of $2\arcsec$.
The beam size of $0.96\arcsec \times 0.82\arcsec$ (PA=$-84.7\arcdeg$) is shown on the bottom-left corner,
corresponding to $21\times 18\,\rm pc^2$ at 4.5~Mpc.
\label{fig:maps}}
\end{figure*}

\begin{figure*}[ht!]
\epsscale{1.15}
\plotone{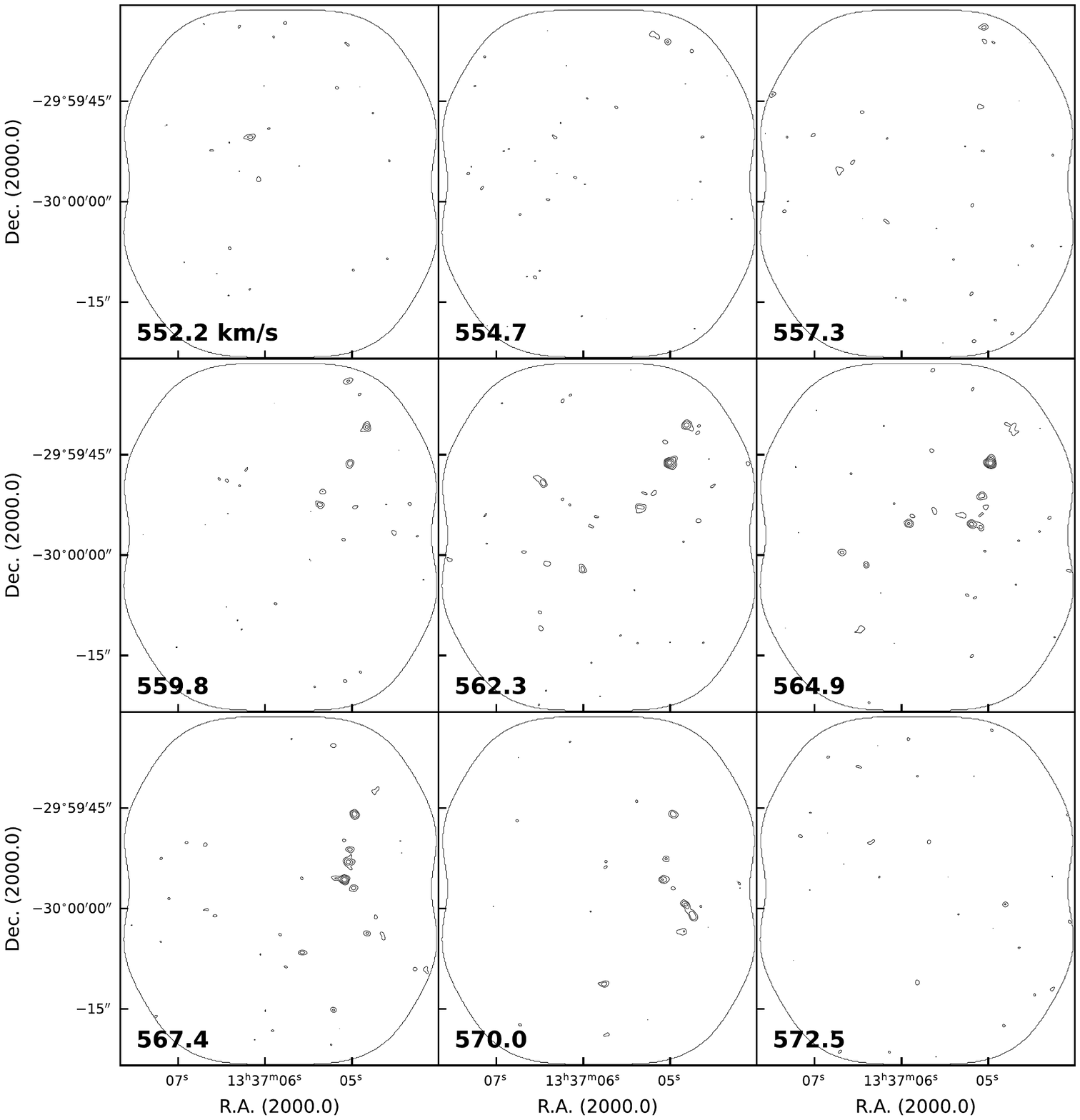}
\caption{
CO(3-2) channel maps with the field of view of the ALMA observations outlined.
Channel velocities are shown on the bottom-left corner.
The RMS noise is 0.96~mJy/beam in the $2.54\,\kmps$ channel,
and the contours are at 3, 5, 8, 12, and 17$\sigma$ significance.
\label{fig:chmap32cont}}
\end{figure*}

\subsection{Supplementary Data}
We obtained the HI 21cm data with the Green Bank Telescope (GBT; project 20A-432) and
combined it with the archival Karl G. Jansky Very Large Array (VLA) data (projects 13B-194 and 14B-192).
We reduced the data in the standard way with GBTIDL and CASA, respectively.
The GBT and VLA data are combined after imaging with CASA's feather task.
The beam size of the GBT+VLA data is $15\arcsec$.
The details of the observations and data reduction will be presented in a separate publication.

The H$\alpha$ narrow-band image was obtained
from narrow (NA656) and broad (Rc) band photometry
with the Subaru Prime Focus Camera (Suprime-Cam) on the Subaru telescope.
The seeing was about $1\arcsec$.
The observations and data reduction were similar to those presented in \citet{Koda:2012ab},
who used the privately-owned old H$\alpha$ filter (NA659).
M83 was re-observed with the observatory-owned filter in the context of a separate, larger H$\alpha$ survey of galaxies for consistency.
The photometric zero points were calibrated using the Pan-STARRS1 Surveys (PS1) Data Release 1 catalog \citep{Chambers:2016aa}.
We followed the procedure described in \citet{Yagi:2013aa} and converted from the PS1 to the Suprime-Cam photometric system
using stellar photometry.
The flux errors are about $10\%$ or less,
and the old and new measurements are consistent within the errors.
Note that discussions in this paper require a flux accuracy of only a factor of $\sim2$.
The details of the H$\alpha$ data will be presented in a separate paper.

\section{Auxiliary Information}

\subsection{HII Regions}\label{sec:Ha}

A classification of HII regions is not within the scope of this paper,
but for the purpose of rough evaluation of SF activity in the observed region,
we visually identified significant H$\alpha$ peaks in the Subaru H$_{\alpha}$ image.
Table \ref{tab:HIIregions} lists the ID, coordinates, aperture radius $r_{\rm ap}$,
and H$\alpha$ luminosity $L_{\rm H_{\alpha}}$.
Figures \ref{fig:fov}e and \ref{fig:fov}f show their locations.
The $L_{\rm H_{\alpha}}$ ranges
from $1\times 10^{35}$ to $3\times 10^{37}\,\rm erg\,s^{-1}$.
\footnote{$L_{\rm H_{\alpha}}$ is from the difference between the narrow and broad band photometry.
Two corrections are accounted for:
other line emission in the narrow band
assuming $\rm [NII]/H\alpha=0.2$, $\rm [SII]/H\alpha=0.12$, and $\rm [OI]/H\alpha=0.03$
from the measurements by \citet{Gil-de-Paz:2007eu} and 
H$\alpha$ emission in the broad band \citep{Yagi:2017aa}.
It turns out that the overall impact of the two corrections is only about 9\% reduction in $L_{\rm H_{\alpha}}$.}
We do not apply an extinction correction
since it is expected to be small \citep{Gil-de-Paz:2007eu} and
is negligible for the discussion here which requires only order of magnitude estimations.
Our HII regions (B, F, J) correspond to (\#09, \#08, \#07) in \citet{Gil-de-Paz:2007eu}
and (\#39, \#41, \#40) in \citet{Bresolin:2009ce}.
The total H$\alpha$ luminosity in our $1\,\kpc^2$ region is $\sim 6.8\times 10^{37}\,\rm erg\,s^{-1}$.
The identified HII regions contain about 80\% of the total luminosity.

As a reference, a model calculation by \citet{Sternberg:2003yb}
suggests that a single B0-type star can produce an HII region of
$L_{\rm H_{\alpha}}=1.4\times 10^{36}\,\rm erg\,s^{-1}$
assuming the Case B recombination and electron temperature of $10^4\,\rm K$.
The HII regions brighter than this likely have O-type stars,
but they have to be relatively late O-type,
as an HII region around a single O6-type star can already
be as bright as $2.7\times 10^{37}\,\rm erg\,s^{-1}$,
similar to the brightest in our XUV region.
As we see below, the observed XUV region has SF activity
similar to, or slightly less than, that of the Orion Nebula,
a relatively minor OB star forming region in the Milky Way.

\begin{deluxetable}{ccccc}
\tablecaption{Visually-Identified HII regions\label{tab:HIIregions}}
\tablehead{
\colhead{ID} & \colhead{RA} &\colhead{DEC} &\colhead{$r_{\rm ap}$} &\colhead{$L_{\rm H_{\alpha}}$} \\
\nocolhead{} & \colhead{(J2000)} & \colhead{(J2000)} & \colhead{($\arcsec$)} & \colhead{($\rm erg\,s^{-1}$)} }
\startdata
A  & 13:37:05.23 & -29:59:33.2 & 2.0 &  $5 \times 10^{36}$ \\
B  & 13:37:04.98 & -29:59:45.9 & 3.1 &  $3 \times 10^{37}$ \\
C  & 13:37:06.41 & -29:59:49.3 & 2.8 &  $2 \times 10^{36}$ \\
D  & 13:37:05.13 & -29:59:50.4 & 1.4 &  $3 \times 10^{35}$ \\
E  & 13:37:05.31 & -29:59:50.6 & 1.1 &  $1 \times 10^{35}$ \\
F  & 13:37:06.60 & -29:59:56.4 & 4.7 &  $1 \times 10^{37}$ \\
G  & 13:37:05.06 & -29:59:58.5 & 2.7 &  $9 \times 10^{35}$ \\
H  & 13:37:05.94 & -30:00:02.9 & 1.9 &  $1 \times 10^{36}$ \\
I  & 13:37:06.27 & -30:00:04.5 & 2.2 &  $1 \times 10^{36}$ \\
J  & 13:37:05.81 & -30:00:11.3 & 2.1 &  $5 \times 10^{36}$ \\
K  & 13:37:05.14 & -30:00:14.5 & 2.4 &  $3 \times 10^{36}$
\enddata
\tablecomments{$L_{\rm H_{\alpha}}$ is a H$_{\alpha}$ luminosity
within an aperture of the radius $r_{\rm ap}$ which is set by visual judgement (see Figure \ref{fig:fov}e).}
\end{deluxetable}

\subsection{The Orion A Molecular Cloud in the MW}\label{sec:oriona}

The measured $L_{\rm H_{\alpha}}$ are similar to those of the Galactic HII regions,
M42 and M43.
Both of them are in the Orion Nebula, which is located in the OMC-1 dense gas clump in the Orion A molecular cloud \footnote{We use the nomenclature for Galactic molecular clouds to characterize their nested internal structures. Clumps are dense gas concentrations ($\gtrsim 10^3\,\rm cm^{-3}$) within clouds ($\gtrsim 10^2\,\rm cm^{-3}$).
See \citet{Bergin:2007aa} for details.}.
They show $L_{\rm H_{\alpha}}=7.1\times 10^{36}$ and
$3.5\times 10^{35}\,\rm erg\,s^{-1}$, respectively (\citealt{Scoville:2001qm}, from radio continuum observations by \citealt{Schraml:1969vi}).
M42 is ionized predominantly by a couple of late O-type stars
(O7 \& O9) with small contributions from a few B-type stars \citep{Hillenbrand:1997vb},
and M43 is excited by a single B0.5-type star \citep{Simon-Diaz:2011wv}.
Hence, individual HII regions in the XUV disk have
similar SF activity to that of the Orion Nebula.
There are also numerous low-mass stars forming across the Orion A cloud \citep{Grossschedl:2019vo},
while their counterparts are not easily detected at the distance of M83.

The Orion A molecular cloud is one of the nearest clouds with massive SF
and arguably the most studied high-mass star forming region.
Its internal structure, i.e, a dense gas clump surrounded by a large volume of
bulk molecular gas, is common among Galactic molecular clouds;
not only among high-mass star forming clouds \cite[e.g., W49 and W51; ][]{Nakamura:1984wa, Watanabe:2017tc, Barnes:2020te},
but also among low-mass star forming clouds \cite[e.g., Perseus, Serpens, and Ophiuchus; ][]{Enoch:2008sv}.
Therefore, we use Orion A as our reference cloud.
If the Orion A cloud (and Orion Nebula) were in the XUV disk of M83,
it would appear as one of brightest UV and H$_{\alpha}$ blobs.
As a gauge of the activity,
compared to other Galactic OB associations, Orion A is ``not particularly
impressive as a massive star-forming region" \citep{Hillenbrand:1997vb}.

The most recent estimation of bulk molecular gas mass in Orion A
is $M_{\rm cloud}=4.0\times 10^4\Msun$,
and its dense clump, OMC-1, has a clump mass of $M_{\rm clump}=7.4\times 10^3\Msun$
\citep{Nakamura:2019wt}.
Thus, the OMC-1 clump contains about 20\% of the cloud mass.
We adopt these masses for our discusson, but note
that an often-quoted older value of $M_{\rm cloud}$ is about two times larger and
is $M_{\rm cloud}=1.0\times 10^5\Msun$ after correction for the distance of 414~pc \citep{Dame:1987aa}.

The star formation rate (SFR) within Orion A is estimated
with several tracer emissions by \citet{Pabst:2021ub} (their Table 5).
For comparisons to the XUV disk later,
the most relevant is SFR$\sim 7.7\times 10^{-5}\,\rm \Msun\, yr^{-1}$, based on H$\alpha$ and $24\mu m$ emission.
The gas consumption timescale by the SF is $M_{\rm cloud}/{\rm SFR}\sim 520\rm\, Myr$.
A caveat is that the SFR estimation in a single cloud is quite uncertain;
with the tracers closely related to high-mass SF
(H$\alpha$, $24\mu m$, far and total infrared emissions),
\citet{Pabst:2021ub} found a range of SFR=$1.6\times 10^{-5}$ to $1.0\times 10^{-4}\,\rm \Msun\, yr^{-1}$.
Keeping this uncertainty in mind,
we will compare SFR in the XUV disk from H$\alpha$ on an assumption of negligible extinction,
and that in Orion A from H$\alpha$ and $24\mu m$ to take extinction into account.
Both of these trace the amount of UV photons from recently-formed massive stars.

\section{Results}\label{sec:results}

Figure \ref{fig:maps} shows (a) the CO(3-2) integrated intensity map,
(b) CO(3-2) peak intensity map (i.e., maximum value along velocity channels at each spatial pixel),
and (c) Band~7 (349~GHz) continuum map.
Only the channels with $>3\sigma$ emission are included for the calculation of (a).
Figure \ref{fig:chmap32cont} shows channel maps.
These maps visually display significant detections of the CO(3-2) emission.
The continuum map shows no significant emission except two potential $3.9\sigma$
detections at (RA,DEC)=(13:37:06.7, -30:00:01.9) and (13:37:05.4, -30:00:08.4),
which do not coincide with the CO(3-2) emission.
Therefore, we discuss mainly about the CO(3-2) emission.

\subsection{Cloud Identification and Parameters}

We identify 23 clouds in the CO(3-2) data in the following way
and list them in Table \ref{tab:clouds}. In the data cube, 
we first identify pixels with $>5\sigma$ peaks and find their envelopes
by expanding their volumes to adjacent pixels down to $3\sigma$ significance.
In some cases the edges of the $3\sigma$ envelopes of two peaks (two clouds)
are connected.
We use the \textsc{CLUMPFIND} algorithm \citep{Williams:1994vg} to split them
-- the partition of the two clouds is determined simply by assigning each pixel to the nearest of the two peaks.

\begin{figure*}[h]
\epsscale{1.0}
\plotone{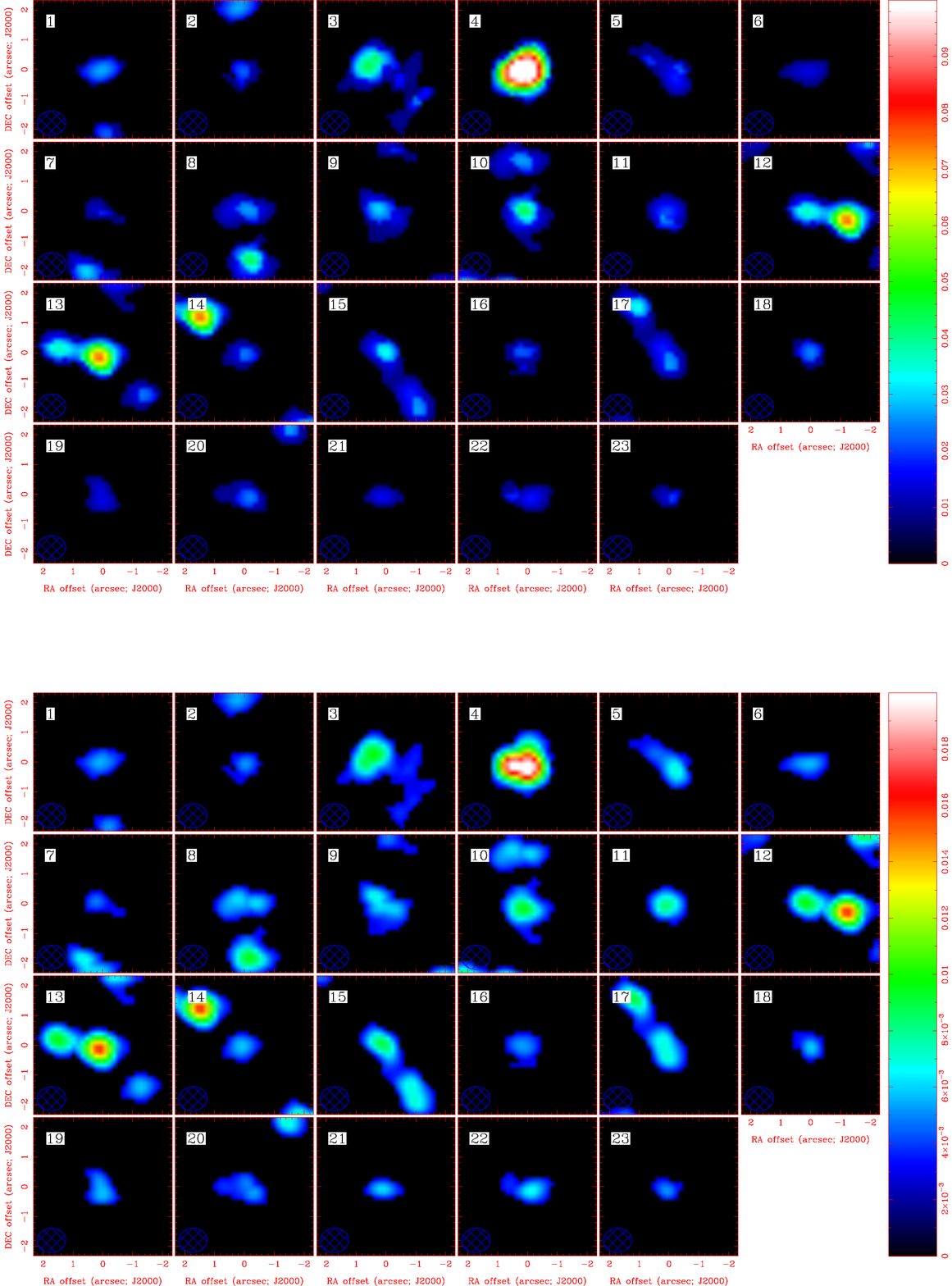}
\caption{
Postage stamp images of individual clouds.
Only the pixels with $>3\sigma$ emission in the data cube are used for these images.
The peaks at the center of each image are the clouds of interest. Surrounding clouds, when they exist, are also in the images.
Top: integrated intensity maps in units of $\rm Jy/beam\,\kmps$.
Bottom: peak intensity maps in units of $\rm Jy/beam$.
The hatched ellipses at the botom-left corners indicate the beam size.
\label{fig:stamps}}
\end{figure*}

\begin{figure*}[h]
\epsscale{1.0}
\plotone{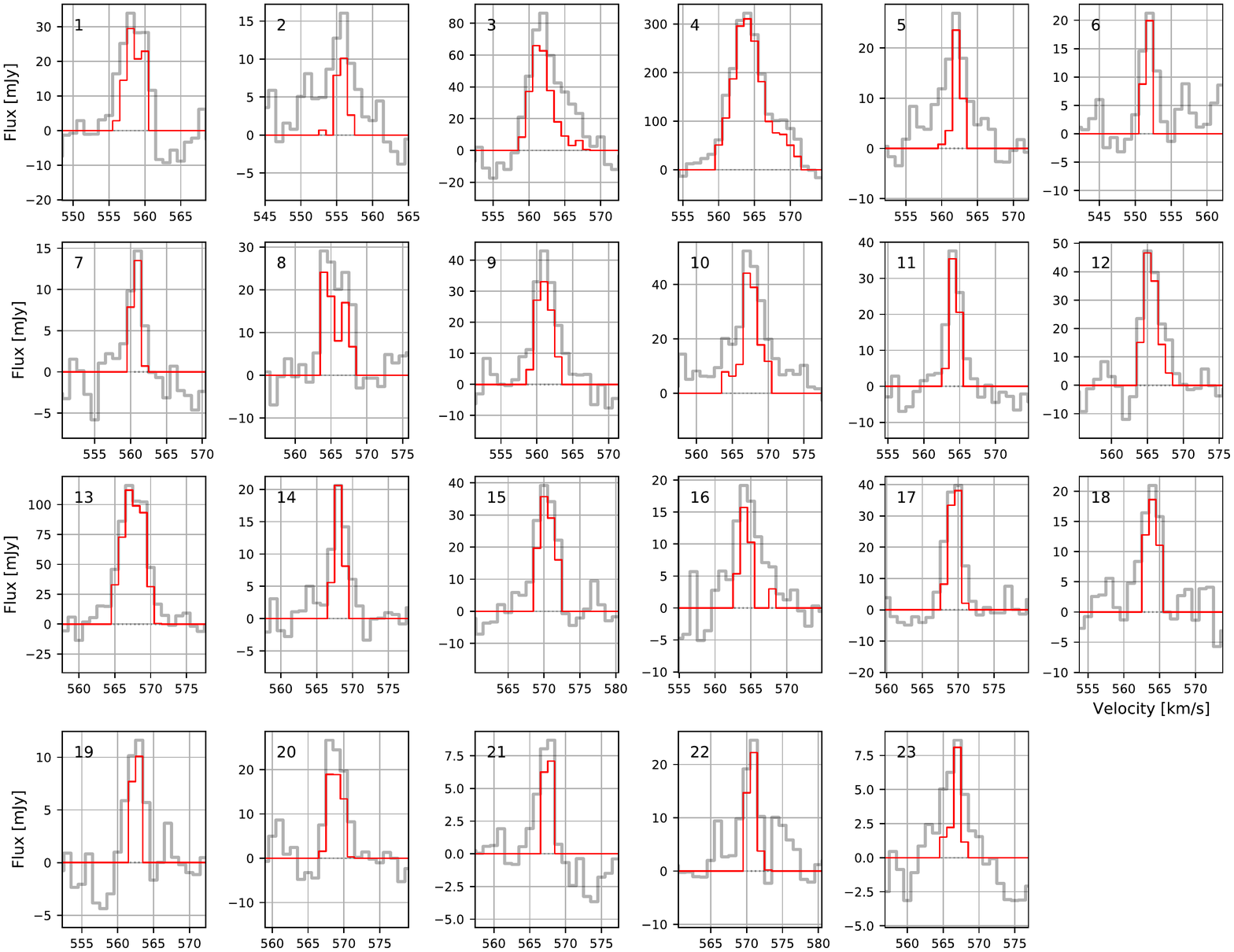}
\caption{
CO(3-2) spectra of the clouds produced with the $1\kmps$ cube.
The spatial pixels and volume that each cloud occupies in the (RA, DEC, Vel) space
are determined with the $2.54\kmps$ cube.
The gray spectra show the sum over all the spatial pixels including noises in each $1\kmps$ channel.
The red spectra include only the pixels with $>3\sigma$ emission in the cloud volume.
\label{fig:spec}}
\end{figure*}

We visually inspect the results and manually merge them back to a single cloud
when the intensity-weighted positions of the split clouds are closer than
the beam size ($0.96\arcsec$) and one velocity pixel ($2.54\kmps$).
Clouds 4, 8, and 20 are merged back.
Cloud 4 might consist of two clouds based on visual inspection,
but we strictly apply the same merge criteria for all cases.
Figure \ref{fig:maps}d shows the locations of the identified clouds
with 2"-diameter circles (about double the beam size).
Figure \ref{fig:stamps} shows the integrated intensity and peak intensity maps of individual clouds.
Figure \ref{fig:spec} shows the spectra of the clouds.
The previous CO(2-1) study did not detect these clouds \citep{Bicalho:2019aa}.
In fact, none of them show $>3\sigma$ emission in the CO(2-1) cube (see Section \ref{sec:mass21}).

Table \ref{tab:clouds} lists the cloud properties.
The first three columns are their intensity-weighted coordinates in RA, DEC, and recession velocity $V$.
The FWHMx and FWHMy are the full width at half maximums
in the RA (=x) and DEC (=y) directions and are
calculated from intensity-weighted dispersions $\sigma_{\rm X}$ (where $X=$x, y).
The $\sigma_{\rm X}$s are calculated without any model profile,
but we convert them to FWHM by assuming a Gaussian distribution
(i.e., ${\rm FWHM}=\sqrt{8\ln 2} \sigma_{\rm X}$),
so that they can be directly compared to the beam size in FWHM.
We note the major axis of the beam is roughly in the RA (x) direction,
and the minor axis is in the DEC (y) direction.
$N_{\rm xy}$ is the number of spatial pixels with $>3\sigma$ emission in the sky projection.
The beam area amounts to $N_{\rm xy}=40$ in FWHM.
$\sigma_{\rm v}$ is the intensity-weighted velocity dispersion calculated with the $1\kmps$ cube.
Only the pixels with $>3\sigma$ emission in the $1\kmps$ cube (red in Figure \ref{fig:spec})
are included for this calculation.
When the number of $>3\sigma$ pixels is less than four,
we place only an upper limit in $\sigma_{\rm v}$, assuming the upper limit of
the FWHM width to be $4\kmps$ (i.e., $\sigma_{\rm v}=1.7\kmps$).
We do not apply deconvolution for the $1\kmps$ channel width
since it is negligible
(i.e., the second moment of the $1\kmps$ boxcar comes to only $\sigma=1/\sqrt{12}\sim 0.29\kmps$,
and its quadratic subtraction from $\sigma_{\rm v}$ makes only a negligible difference).
The peak intensity, $S_{\rm CO(3-2)}^{\rm peak}$, and total flux, $S_{\rm CO(3-2)}dv$, are also listed in the table.
The error in total flux is calculated by shifting the cloud volume within
the 3D data cube in random directions 300 times and
by calculating the standard deviation among the shifted volumes.

\subsection{Sizes} \label{sec:size}

Some clouds apparently show internal structures in Figure \ref{fig:stamps}
and are on the verge of being spatially resolved.
On the other hand, most have FWHM sizes similar to, or smaller than, the beam size.
Thus, all clouds, except Cloud 4 and 8, are spatially unresolved
or only very marginally resolved
at the resolution of $0.96\arcsec$ (21 pc).
Clouds 3, 5, and 22 show FWHMx and/or FHWMy larger than the beam size (Table \ref{tab:clouds}),
but they suffer from low level emission (Figure \ref{fig:stamps}).
For example, Cloud 3 has a large peculiar extension
at around 3-4$\sigma$ significance,
which increases FWHMx and FHWMy, but appears spurious.
The extended components of Cloud 5 and 22 are also at low level (3-4$\sigma$).
We also note that FWHMy is measured in about the direction of the beam minor axis.
It is often smaller than the minor axis size of $0.82\arcsec$,
again because the measurements suffer from the low level emission.

Cloud 4 and 8 show two peaks in Figure \ref{fig:stamps},
which are merged into a single cloud in our cloud identification process
since they are closer than the resolution.
We consider them as slightly resolved single clouds
with a caveat that they could be a blend of two unresolved clouds.

Even in the marginally-resolved cases, measuring cloud sizes is difficult.
The cloud radius $R$ may be estimated with the equation $R=1.92\sigma_{\rm r}$ \citep{Solomon:1987pr}
using the beam-deconvolved average standard deviation in radius
\begin{equation}
   \sigma_{\rm r}= \sqrt{ \frac{{\rm FWHMx} {\rm FWHMy}-b_{\rm maj}b_{\rm min}}{8\ln 2} }.
\end{equation}
Since the measured size is sometimes smaller than the beam size (Table \ref{tab:clouds}),
we consider $\sigma_{\rm r}$ and $R$ measurable only when the measured size is at least 20\% larger than the beam size
(assuming that FWHMx and FWHMy are measured with 20\% accuracy).
This sets the minimum measurable $R$ to be $R=0.54\sqrt{b_{\rm maj}b_{\rm min}}=0.48\arcsec$ ($10.5\pc$).
Otherwise, we have only an upper limit of $10.5\pc$.
In this way, only one cloud with the spurious extension (Cloud 3; Figure \ref{fig:stamps}) gives a measurable $R$,
while, as discussed, it suffers from the error.
Therefore, the CO(3-2)-emitting parts in the detected clouds (their diameters) are smaller than the resolution of 21~pc.

As a reference, the Orion A cloud
has a whole extent of $\sim 20\pc$ when it is observed in the CO(1-0) or CO(2-1) emission
which trace the bulk molecular gas \citep{Sakamoto:1994lr, Nakamura:2019wt, Kong:2018aa}.
Its star-forming dense clump OMC-1
occupies only a small area  ($\sim 2\pc$) at the heart of the cloud
and is bright in CO(3-2) \citep{Ikeda:1999vt}.
It is possible that our detected CO(3-2) peaks are
mainly from dense clumps within molecular clouds,
and hence, are unresolved.

\subsection{Velocity Widths} \label{sec:veldisp}

The velocity widths of most clouds are resolved in the $1\kmps$ cube
(see $\sigma_{\rm v}$ in Table \ref{tab:clouds}).
Cloud 3 has a relatively large $\sigma_{\rm v}$ due to the spurious extension.
Cloud 4 has a wide width of $\sigma_{\rm v}=6.1\kmps$ with a caveat that it could be a blend of two clouds.

\begin{figure}[h]
\epsscale{1.15}
\plotone{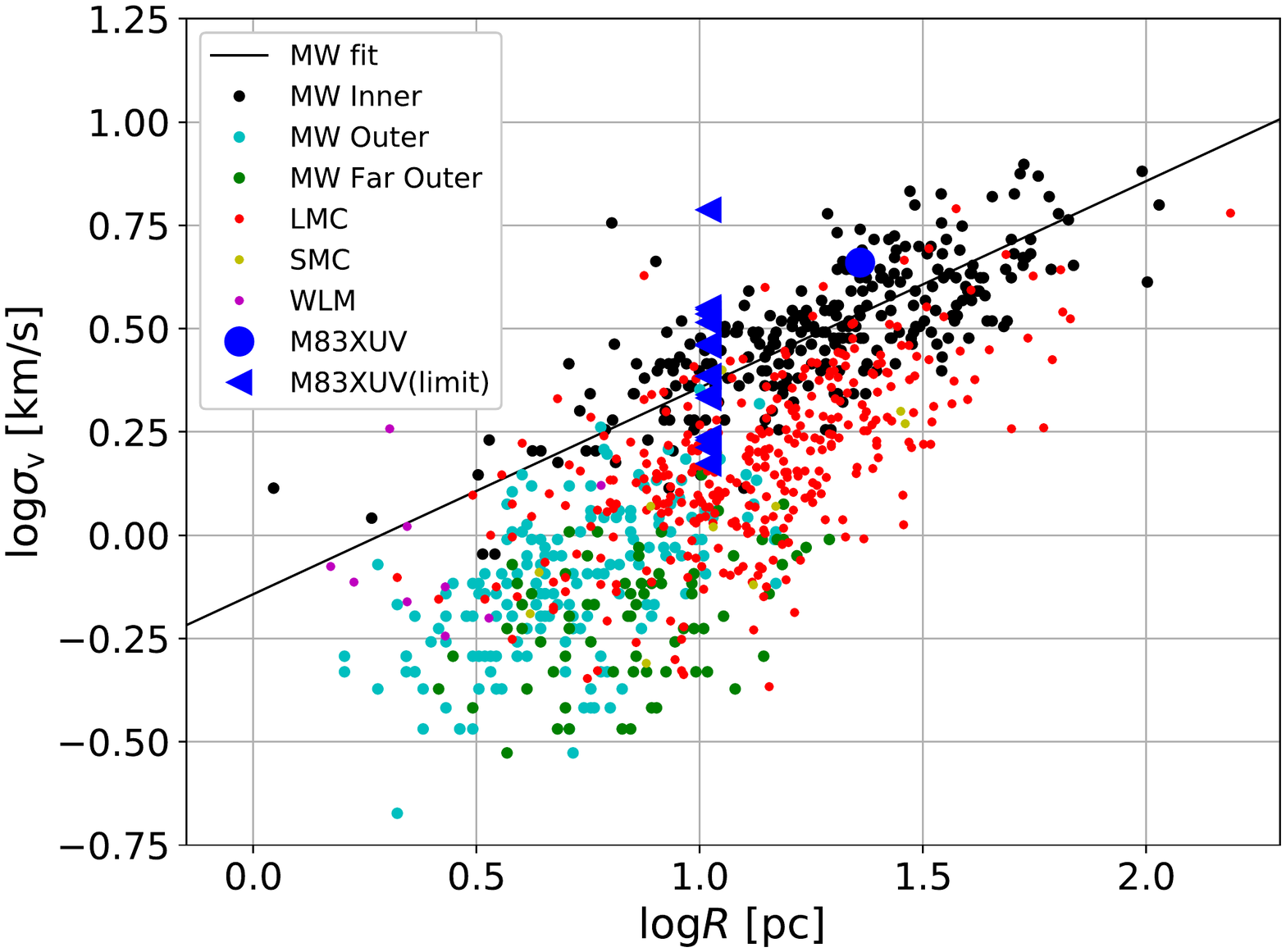}
\caption{
Cloud radius-velocity dispersion plot.
M83 XUV clouds are blue (and the blue arrows are the upper limits in radius, while they are resolved in velocity).
Reference points in the background are for the MW's inner disk \citep{Solomon:1987pr},
outer disk \citep[$r_{\rm gal}\sim$12-14$\kpc$; ][]{Sun:2017vv},
and far outer disk \citep[$\sim$14-22$\kpc$; ][]{Sun:2015wr}.
Large and Small Magellanic Clouds \citep{Wong:2011td, Bolatto:2008ul},
and WLM \citep{Rubio:2015vc}.
The solid line is a fit to the inner disk clouds \citep{Solomon:1987pr}.
\label{fig:scaling}}
\end{figure}

Again using Orion A as a reference, the velocity width of the bulk gas, measured in CO(1-0)
and averaged over the whole cloud,
is about $5\kmps$ at 50\% of the peak brightness \citep[from Figure 5 of ][]{Nakamura:2019wt}.
The full CO(3-2) velocity width over the OMC-1 clump (size of about $2\pc$)
is not found in the literature;
toward Orion KL (the brightest region), it is about $7\kmps$ at 50\% 
of the peak brightness in a $134\arcsec$ ($\sim 0.3\pc$) beam \cite[from Figure 15 of][]{Masui:2021tj}.
Hence, the dense part has a broader linewidth than the entire cloud.
These are similar to the velocity widths of Cloud 4 measured in CO(3-2).

Figure \ref{fig:scaling} plots $\sigma_{\rm v}$ against the radius $R$ (mostly the upper limits; see Section \ref{sec:size}).
It also shows clouds in some other galaxies \citep{Solomon:1987pr, Bolatto:2008ul, Wong:2011td, Rubio:2015vc},
with notes that they are measured in CO(1-0) or CO(2-1) while our XUV clouds are in CO(3-2)
and that the different CO transitions may trace different parts of clouds (Section \ref{sec:clstructures}).
For their measured $\sigma_{\rm v}$,
our XUV clouds generally have smaller (or similar) sizes compared to the clouds in the other galaxies.
This makes sense if the CO(3-2) emission is primarily tracing dense clumps within the clouds, as seen in Orion A.

\subsection{Distribution}

The 23 clouds are distributed over a large area across the $1\kpc^2$
field of view (Figures \ref{fig:maps}d and \ref{fig:chmap32cont}).
Their overall distribution shows some coherent (not random) structures,
which appear like chains of clouds.
These chains extend over 600~pc in length,
or potentially longer beyond the field of view.
The cloud distribution also shows a large-scale velocity gradient
in the channel maps (Figure \ref{fig:chmap32cont}).
From the low to high velocity channels, it spatially shifts from the northeast to the southwest side.

\begin{figure*}[h]
\epsscale{1.15}
\plotone{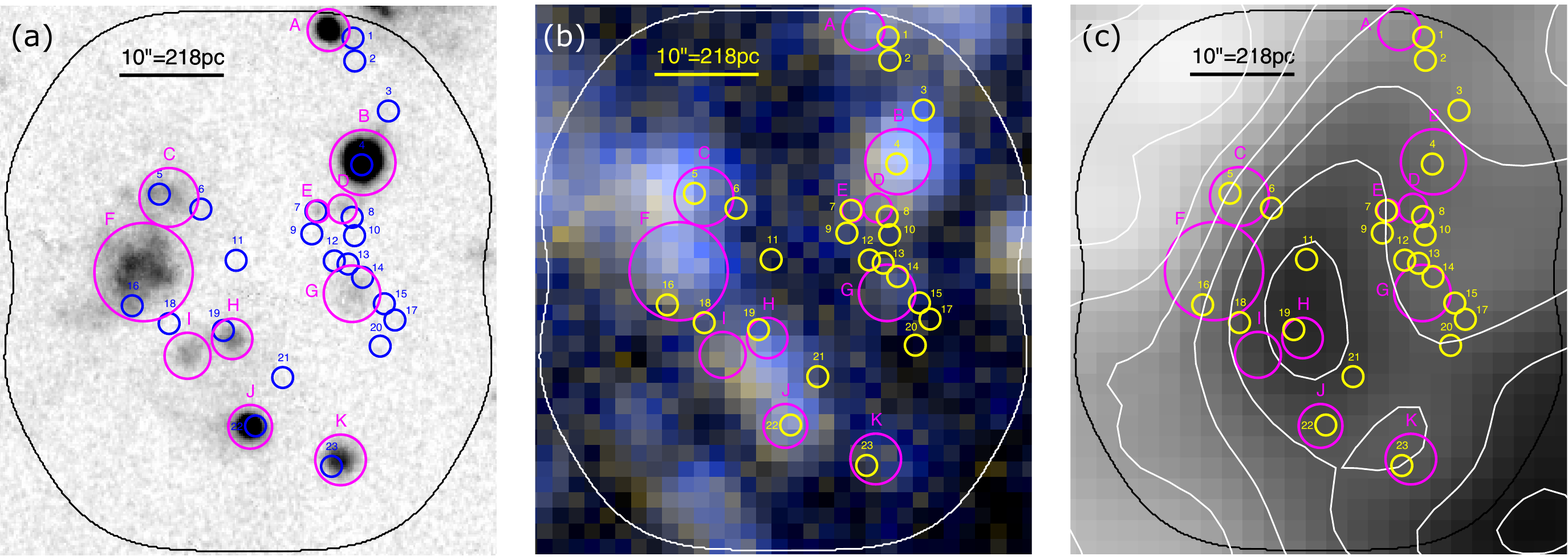}
\caption{
(a) Subaru H$\alpha$ image at a resolution of about $1\arcsec$.
(b) GALEX FUV \& NUV-band color composite image at a resolution of about $5\arcsec$ \citep{Gil-de-Paz:2007lj}.
(c) VLA+GBT HI 21cm image (grayscale) with contours -- from the brightest -- at 0.30, 0.25, 0.20, 0.15, 0.10, and 0.05 ${\rm Jy/beam}\,\kmps$ with the $15\arcsec$ beam. The brightest contour corresponds to the HI surface density of $\sim 12 \,\Msun\,\pc^{-2}$ (and $16 \,\Msun\,\pc^{-2}$ for HI+He).
Magenta circles denote the HII regions identified by eye with the labels of 'A', 'B', 'C', ...
The H$\alpha$ luminosities $L_{\rm H_{\alpha}}$ enclosed within each circle is listed in Table \ref{tab:HIIregions}.
Blue or yellow circles are molecular clouds and are the same as the ones in Figure \ref{fig:maps}d with the labels of '1', '2', '3', ...
\label{fig:UVHaCO}}
\end{figure*}

Figure \ref{fig:UVHaCO} shows the cloud locations on the H$\alpha$, UV, and HI images.
Cloud 4, 7, 22, 23, and possibly 5, 16, 19, are directly associated with HII regions B, E, J, K, C, F, and H, respectively.
Among these, B, J, K, and C are four of the most prominent HII regions in the field.
Cloud 4 is the brightest in CO(3-2),
while, interestingly, the other three clouds are in the bottom 2/3 of the clouds in total CO(3-2) flux.
All the other clouds are in the proximity of, but outside, the HII regions.
The variations could be due to an evolutionary sequence of molecular clouds,
or to the stochasticity of massive SF in this low SF environment
\citep[i.e., when the total stellar mass produced in a region is low, clouds may not always produce
stars over the full mass spectrum. The stellar initial mass function may be filled stochastically,
but not completely; ][]{Koda:2012ab}.
With respect to the atomic gas distribution,
most clouds and HII regions are around, but not on, the peak of the HI emission.

\floattable
\begin{deluxetable}{cccccccccccc}
\tablecaption{Cloud Parameters\label{tab:clouds}}
\tablehead{
\colhead{(1)} & \colhead{(2)}  & \colhead{(3)} & \colhead{(4)} & \colhead{(5)} & \colhead{(6)} & \colhead{(7)} & \colhead{(8)} & \colhead{(9)}  & \colhead{(10)} & \colhead{(11)} & \colhead{(12)}\\
\cline{1-12}
\colhead{ID} & \colhead{RA}  & \colhead{DEC} & \colhead{$V$}& \colhead{FWHMx} & \colhead{FWHMy}   & \colhead{$N_{\rm xy}$}  & \colhead{$\sigma_{\rm v}$} & \colhead{$S_{\rm CO(3-2)}^{\rm peak}$} & \colhead{$S_{\rm CO(3-2)}dv$} & \colhead{$M_{\rm clump}$} & \colhead{$M_{\rm cloud}$} \\
\nocolhead{} & \colhead{(J2000)}  & \colhead{(J2000)} & \colhead{($\kmps$)}  & \colhead{(")}   & \colhead{(")}  & \nocolhead{} & \colhead{($\kmps$)} & \colhead{(mJy/beam)} & \colhead{($\rm mJy \kmps$)} & \colhead{($\Msun$)} & \colhead{($\Msun$)}
}
\startdata
1 & 13:37:05.05 & -29:59:34.0 & 558.5 & 0.81 & 0.52 & 52 & 2.9 & 5.59 & $22.6\pm 2.8$ & $8.1\times 10^{2}$ & $2.5\times 10^{3}$  \\
2 & 13:37:05.04 & -29:59:36.2 & 555.0 & 0.59 & 0.57 & 39 & 2.1 & 5.30 & $12.6\pm 1.7$ & $4.6\times 10^{2}$ & $1.4\times 10^{3}$  \\
3 & 13:37:04.79 & -29:59:41.0 & 562.5 & 1.67 & 1.46 & 197 & 4.6 & 9.83 & $79.2\pm 4.9$ & $2.9\times 10^{3}$ & $8.7\times 10^{3}$  \\
4 & 13:37:04.99 & -29:59:46.1 & 564.4 & 1.02 & 0.95 & 156 & 6.1 & 21.68 & $207.7\pm 7.3$ & $7.5\times 10^{3}$ & $2.3\times 10^{4}$  \\
5 & 13:37:06.48 & -29:59:49.0 & 562.1 & 1.13 & 0.98 & 85 & 1.5 & 7.08 & $25.5\pm 2.6$ & $9.2\times 10^{2}$ & $2.8\times 10^{3}$  \\
6 & 13:37:06.17 & -29:59:50.4 & 552.2 & 0.93 & 0.50 & 51 & $<$1.7 & 5.77 & $13.4\pm 2.0$ & $4.8\times 10^{2}$ & $1.5\times 10^{3}$  \\
7 & 13:37:05.32 & -29:59:50.6 & 560.5 & 0.78 & 0.48 & 33 & $<$1.7 & 4.94 & $8.1\pm 1.4$ & $2.9\times 10^{2}$ & $8.9\times 10^{2}$  \\
8 & 13:37:05.06 & -29:59:51.2 & 565.8 & 1.13 & 0.60 & 83 & 3.6 & 6.21 & $28.3\pm 3.0$ & $1.0\times 10^{3}$ & $3.1\times 10^{3}$  \\
9 & 13:37:05.35 & -29:59:52.8 & 561.4 & 0.93 & 0.89 & 96 & 2.9 & 7.06 & $33.8\pm 3.3$ & $1.2\times 10^{3}$ & $3.7\times 10^{3}$  \\
10 & 13:37:05.04 & -29:59:53.0 & 567.5 & 0.82 & 0.95 & 101 & 3.3 & 9.55 & $43.4\pm 3.5$ & $1.6\times 10^{3}$ & $4.8\times 10^{3}$  \\
11 & 13:37:05.91 & -29:59:55.3 & 564.6 & 0.73 & 0.70 & 62 & $<$1.7 & 8.39 & $22.2\pm 2.3$ & $8.0\times 10^{2}$ & $2.4\times 10^{3}$  \\
12 & 13:37:05.19 & -29:59:55.4 & 565.5 & 0.77 & 0.60 & 61 & 2.4 & 9.73 & $30.7\pm 2.9$ & $1.1\times 10^{3}$ & $3.4\times 10^{3}$  \\
13 & 13:37:05.09 & -29:59:55.7 & 567.7 & 0.84 & 0.71 & 89 & 3.4 & 15.70 & $74.0\pm 4.2$ & $2.7\times 10^{3}$ & $8.1\times 10^{3}$  \\
14 & 13:37:04.98 & -29:59:57.0 & 567.8 & 0.70 & 0.52 & 45 & $<$1.7 & 5.96 & $14.5\pm 1.9$ & $5.2\times 10^{2}$ & $1.6\times 10^{3}$  \\
15 & 13:37:04.82 & -29:59:59.5 & 570.4 & 0.76 & 0.87 & 75 & 2.4 & 9.01 & $29.2\pm 2.6$ & $1.1\times 10^{3}$ & $3.2\times 10^{3}$  \\
16 & 13:37:06.68 & -29:59:59.7 & 564.9 & 0.65 & 0.67 & 51 & 3.5 & 5.33 & $15.0\pm 2.2$ & $5.4\times 10^{2}$ & $1.6\times 10^{3}$  \\
17 & 13:37:04.75 & -30:00:01.1 & 569.8 & 0.77 & 0.91 & 78 & 1.7 & 7.21 & $27.3\pm 2.7$ & $9.8\times 10^{2}$ & $3.0\times 10^{3}$  \\
18 & 13:37:06.41 & -30:00:01.4 & 563.8 & 0.57 & 0.53 & 38 & $<$1.7 & 6.02 & $14.1\pm 1.9$ & $5.1\times 10^{2}$ & $1.5\times 10^{3}$  \\
19 & 13:37:06.01 & -30:00:02.1 & 562.3 & 0.61 & 0.75 & 46 & $<$1.7 & 5.81 & $12.7\pm 1.7$ & $4.6\times 10^{2}$ & $1.4\times 10^{3}$  \\
20 & 13:37:04.85 & -30:00:03.6 & 569.0 & 0.96 & 0.55 & 61 & 2.2 & 5.43 & $19.2\pm 2.3$ & $6.9\times 10^{2}$ & $2.1\times 10^{3}$  \\
21 & 13:37:05.57 & -30:00:06.6 & 567.4 & 0.72 & 0.41 & 34 & $<$1.7 & 6.30 & $9.4\pm 1.4$ & $3.4\times 10^{2}$ & $1.0\times 10^{3}$  \\
22 & 13:37:05.77 & -30:00:11.2 & 570.5 & 1.02 & 0.53 & 58 & 1.6 & 6.78 & $16.9\pm 1.9$ & $6.1\times 10^{2}$ & $1.9\times 10^{3}$  \\
23 & 13:37:05.21 & -30:00:15.1 & 566.9 & 0.57 & 0.40 & 26 & 1.7 & 5.23 & $7.4\pm 1.3$ & $2.7\times 10^{2}$ & $8.2\times 10^{2}$
\enddata
\tablecomments{
(1) Cloud ID.
(2)(3) Coordinate.
(4) Recession velocity $V$.
(5)(6) FWHM widths in RA (=x) and DEC (=y) directions, calculated from intensity-weighted dispersions and assuming Gaussian profile.
(7) Number of pixels with $>3\sigma$ emission in the sky projection.
(8) Velocity dispersion measured with the $1\kmps$ cube.
(9) Peak intensity within cloud.
(10) Total flux of cloud. The error is a random error and does not include systematic errors.
(11)(12) Clump and cloud masses, including CO-dark H$_2$ as well as He. $M_{\rm cloud}=M_{\rm clump}/0.34$.
Note that the beam-deconvolved radius $R$ is not listed, since it is only an upper limit ($R<0.48\arcsec$) for all clouds except Cloud 3 ($R=1.05\arcsec$) whose measurement suffers from the spurious extension (see Section \ref{sec:size}).
}
\end{deluxetable}

\section{Discussion} \label{sec:clstructures}

\subsection{Molecular Clouds in Low Metallicity}
\begin{figure}[ht!]
\epsscale{1.0}
\plotone{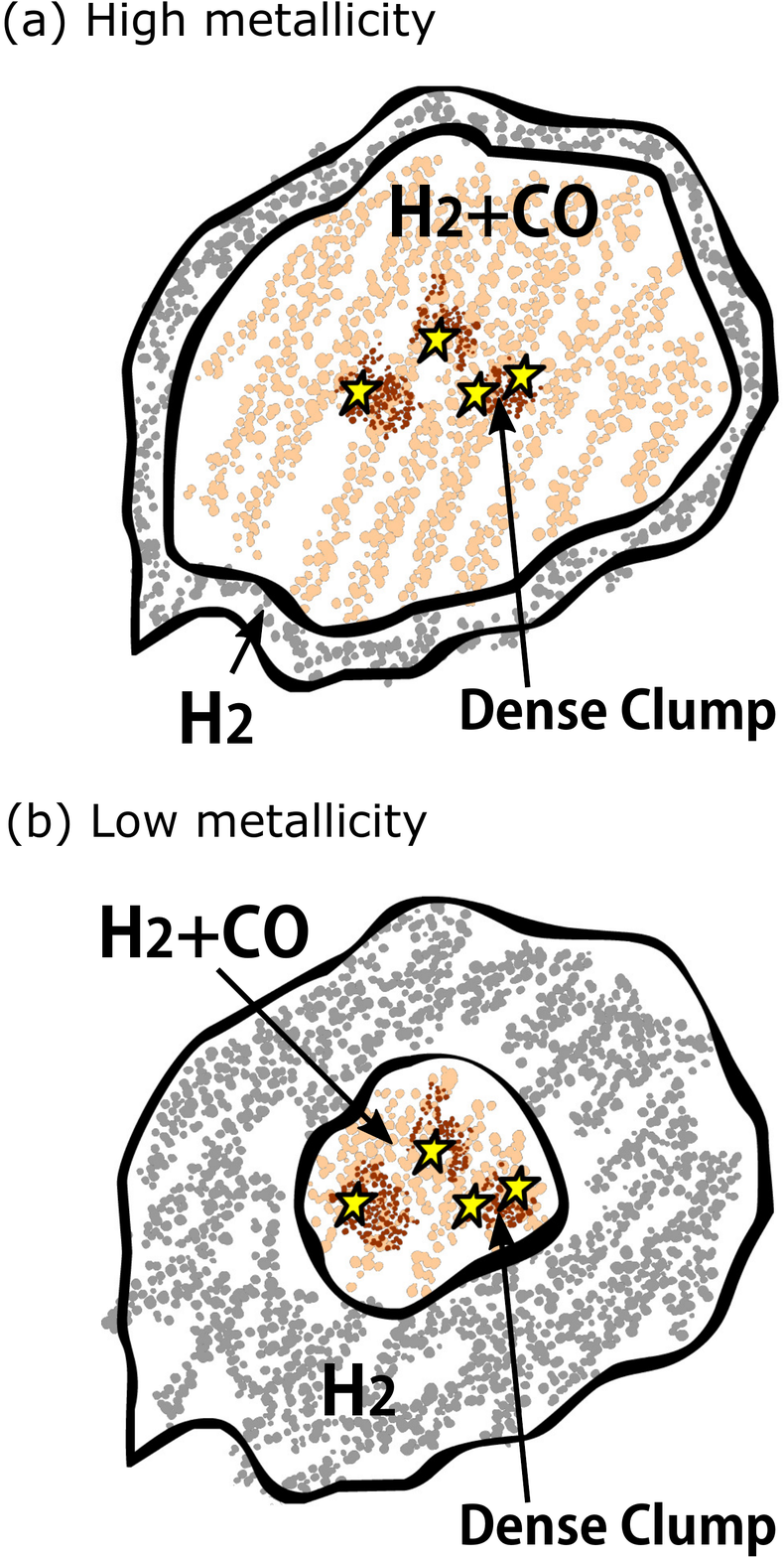}
\caption{
Schematic illustrations of the physical and chemical structures of
a molecular cloud 
(a) in high-metallicity environment, and
(b) in low-metallicity environment.
The physical structures (i.e., mass distribution and SF activity)
are common between (a) and (b), but their chemical structures (i.e., CO abundance) are different.
The ambient radiation penetrates into the cloud,
dissociates CO molecules in the cloud outer layers,
and makes them CO-deficient and CO-dark,
while more abundant H$_2$ molecules can self-shield themselves and remain (gray).
These CO-dark layers thicken at low metallicity,
since a lower dust abundance results in  less efficient extinction and
allows the radiation to penetrate deeper into the cloud  \citep[e.g., ][]{Maloney:1988lr, van-Dishoeck:1988br}.
Still, the CO molecules can remain near the cloud center (orange).
We hypothesize that the dense clumps (brown), often associated with SF (yellow), are located
at the heart of the cloud and likely remain intact with abundant CO molecules.
The CO(3-2) emission originates from the dense clump region.
\label{fig:schematic}}
\end{figure}

We are using the Orion A molecular cloud as a fiducial reference cloud
given the similarity of its SF activity to the (brightest) clouds in the XUV disk.
Orion A has an internal mass distribution common among other large and small Galactic molecular clouds,
having small star-forming dense clump(s) embedded deeply in thick layers of bulk molecular gas
\citep{Nakamura:1984wa, Enoch:2008sv, Watanabe:2017tc, Barnes:2020te}.
Figure \ref{fig:schematic}a schematically illustrates this mass distribution.
In terms of CO emissions, 
CO(1-0) and CO(2-1) can be excited easily at an average density and temperature
in molecular clouds \citep[$\sim 10^2\,\rm cm^{-3}$, $\sim 10\,\rm K$; ][]{Scoville:1987vo},
and hence, the bulk gas contributes significantly to
their total CO(1-0) and CO(2-1) luminosities \citep{Kong:2018aa, Nakamura:2019wt}.
By having the $J=3$ level temperature of $E_{J=3}/k\sim 33.2\,\rm K$ and critical density of $>10^3\,\rm cm^{-3}$,
the CO(3-2) emission requires higher density and temperature than the averages for excitation
and is radiated predominantly from the dense clumps \citep{Ikeda:1999vt}.
Once it is excited in thermalized gas, CO(3-2) can become brighter than CO(1-0) and CO(2-1) in total flux.

We hypothesize that the clouds in the XUV disk share the common clump-envelope mass distribution
of the Galactic molecular clouds.
The common mass structure is expected for gravitationally-bound clouds
as it is determined predominantly by the internal physics
rather than the environment.

Even when the mass distribution is the same,
the chemical structure and appearance of the clouds in the CO emissions
depend on the metal abundance \citep[e.g., ][]{Maloney:1988lr, van-Dishoeck:1988br, Israel:1997lr, Pak:1998vb, Wolfire:2010aa}.
In high metallicity ($\sim 1\Zsun$; Figure \ref{fig:schematic}a),
the bulk gas in clouds can be detected in the low excitation transitions
($J=$1-0 or 2-1) of CO --
the brightest and second-most-abundant molecule after H$_2$.
In low metallicity ($< 1\Zsun$; ; Figure \ref{fig:schematic}b), the dust extinction is low,
and the ambient stellar radiation can penetrate deeper into the clouds.
This radiation dissociates more CO in their outer layers through CO line absorptions,
while the abundant H$_2$ molecules can be self-shielded.
This differential photo-dissociation makes the cloud outer layers CO-deficient and CO-dark
while still H$_2$-abundant (Figure \ref{fig:schematic}b).

The ambient radiation can come from the UV-emitting sources across the $1\kpc^2$ region (Figure \ref{fig:fov}f)
and be potentially as strong as that in the solar neighborhood.
There are several Orion-class SF regions within the region (Table \ref{tab:HIIregions}, Figure \ref{fig:UVHaCO}).
In the solar neighborhood,
the Orion cloud is the closest massive SF region at a distance of 414~pc.

In addition to this photo-dissociation model for the cloud envelopes, we hypothesize that
the dense clumps reside in the self-shielded, CO-abundant part of the clouds.
The clouds in the XUV disk may suffer from this differential photo-dissociation
due to the low metallicity.
With this hypothesis,
we can consistently explain the detection of CO(3-2), which is mainly from the dense clumps;
non-detection of CO(2-1);
as well as H$\alpha$ luminosities, tracing SF activity.

\subsection{Dense Clumps at Hearts of Clouds}\label{sec:lowzstructure}

In our hypothesis, the dense clumps can be shielded from photo-dissociation and
radiate CO emission even in the low-metallicity environment (Figure \ref{fig:schematic}b).
The fractions of the CO-abundant part and dense clumps over entire clouds can be crudely estimated
from a combination of previous measurements of the mass-to-light ratio $M_{\rm cloud}/L_{\rm CO}$
-- between the cloud's total mass $M_{\rm cloud}$ and CO luminosity $L_{\rm CO}$.
The $M_{\rm cloud}/L_{\rm CO}$ is often denoted as
the CO-to-H$_2$ conversion factor $X_{\rm CO}$ and $X_{\rm CO}\propto M_{\rm cloud}/L_{\rm CO}$.
\footnote{In units of $\,\rm H_2 \, cm^{-2} [K\cdot \kmps]^{-1}$
which will be omitted in the rest of the main text.}
The following estimation supports the hypothesis that the dense clumps are indeed protected
even in low metallicity, and thus, can radiate CO(3-2) emission as much as
their counterparts in high metallicity.

The clouds of the same mass $M_{\rm cloud}$ in high and low metallicity have different
amounts and distributions of CO molecules (Figure \ref{fig:schematic}a,b).
The ones in low metallicity show a smaller $L_{\rm CO}$, and hence larger $X_{\rm CO}$.
This difference is detected using the virial mass measurements of molecular clouds with CO(1-0):
$X_{\rm CO(1-0)}=3\times 10^{20}$ in the Milky Way \citep[$\sim 1\Zsun$; ][]{Scoville:2016aa}
and $7\times 10^{20}$ in the LMC \citep[$\sim 0.4\Zsun$; ][]{Fukui:2008tl}.
Therefore, for the same mass, clouds in the LMC emit only about $40\%$ of $L_{\rm CO(1-0)}$
compared to their Galactic counterparts of the same $M_{\rm cloud}$.
In other words, only about $40\%$ of molecular gas, presumably at cloud centers,
would have CO molecules at the LMC metallicity.

The $40\%$ of molecular gas that has CO likely hosts dense clumps at the cloud centers (Figure \ref{fig:schematic}b).
The CO(3-2) emission requires higher density and/or kinetic temperature
for excitation than CO(1-0) and CO(2-1),
and preferentially traces the dense clumps rather than
the bulk gas \citep[e.g., ][]{Komugi:2007un, Wilson:2009fk}. \citet{Wilson:2009fk} compared the CO(3-2) luminosity $L_{\rm CO(3-2)}$ against
the bulk mass $M_{\rm cloud}$ in nearby galaxies at solar metallicity ($\sim 1\Zsun$).
Their result suggests that the detected $L_{\rm CO(3-2)}$ are only 
34\% of those expected from $L_{\rm CO(1-0)}$ on an assumption of thermalized gas
(i.e., the CO 3-2/1-0 ratio in the brightness temperature is about unity).
Thus, only about 34\% of molecular gas in clouds radiates CO(3-2) emission at the high metallicity
(more accurately, the molecular gas that emits about 34\% of $L_{\rm CO(1-0)}$ radiates CO(3-2) emission).
Therefore, at LMC metallicity, we expect the dense clumps emitting at CO(3-2) (34\% of the molecular gas) are located in the same region as the gas that is protected from photo-dissociation and has CO molecules (40\% of the molecular gas), as in Figure \ref{fig:schematic}b.

\citet{Leroy:2022tj} recently found a CO 3-2/1-0 ratio of 0.31 (0.20-0.42 at 16th to 84th percentiles)
as galaxy-integrated mean values among 20 massive solar metallicity galaxies.
This is consistent with our adopted ratio of 0.34 by \citet{Wilson:2009fk}.
Our discussions in this paper are for typical, Orion-class SF regions (Figure \ref{sec:Ha}),
but the ratio may increase mildly in more intense SF regions
\citep{Onodera:2012lr, Miura:2012yq, Vlahakis:2013aa, Miura:2014tq, Morokuma-Matsui:2017ua}.

\subsection{Masses of Molecular Clouds and Clumps} \label{sec:mass}

Even when the cloud outer layers become CO-dark in low metallicity,
there are still abundant H$_2$ molecules there,
and $M_{\rm cloud}$ does not depend on the metallicity.
In our hypothesis, the dense clumps are protected from photo-dissociation and
would radiate a similar amount of $L_{\rm CO(3-2)}$ in high and low metallicities.
Hence, the calibration of $M_{\rm cloud}/L_{\rm CO(3-2)}$ in high metallicity
should be applicable in low metallicity,
and the conversion equation from $L_{\rm CO(3-2)}$ to $M_{\rm cloud}$
derived in high metallicity can be used here.
We adopt the calibration by \citet{Wilson:2009fk},
\begin{eqnarray}
M_{\rm cloud} 
    &=& 1.1 \times 10^5 \Msun \left(\frac{S_{\rm CO(3-2)} dv}{1\,\rm Jy\cdot \kmps}\right)\left(\frac{D}{4.5\rm \, Mpc}\right)^{2} \nonumber \\
    &\times & \left( \frac{\left< R_{3-2/1-0} \right>}{0.34} \right)^{-1} \left(\frac{X_{\rm CO(1-0)}}{3.0\times 10^{20}} \right), \label{eq:mcloud3-2}
\end{eqnarray}
where the CO(3-2) flux $S_{\rm CO(3-2)} dv$ is related to the luminosity
$L_{\rm CO(3-2)}=4\pi D^2 S_{\rm CO(3-2)} dv$.
$\left< R_{3-2/1-0} \right>$ is the CO 3-2/1-0 brightness temperature ratio averaged over galaxies with high metallicity.
The $X_{\rm CO(1-0)}$ term is from the calibration made in the high metallicity,
and hence the $X_{\rm CO(1-0)}$ in high metallicity should be used here
even when it is applied in low metallicity.
$M_{\rm cloud}$ from CO(3-2) is listed in Table \ref{tab:clouds}
and ranges from $8.2\times 10^2\Msun$ to $2.3\times 10^4\Msun$.

The derived masses are much lower than the previous detections
in XUV disks \citep{Braine:2004aa, Braine:2007aa, Braine:2010aa, Braine:2012aa, Dessauges-Zavadsky:2014fk}.
From their high abundance in our region, the low-mass molecular clouds are
likely a more common cloud population in XUV disks than previously detected.
The most massive cloud (Cloud 4) has a mass similar to that of Orion A.
Its SF activity measured in $L_{H\alpha}$ is also similar to that of the Orion Nebula in Orion A (Section \ref{sec:oriona}).
Note again that among massive star-forming regions in the MW,
Orion A is one of the least impressive.
All the other clouds in our study have HII regions in their proximity, but are
less massive and less active in SF.
The least massive cloud of $\sim 800\Msun$, only 4\% of Orion A's mass,
is unlikely to form OB stars.
The SF in the XUV disk appears to be limited by the cloud masses.

The CO(3-2) emission is mainly from dense clumps in the clouds
since it requires high density and temperature for excitation.
We attempt a crude estimation of the clump masses
$M_{\rm clump}$ under a set of assumptions:
(1) the clumps do not suffer from photo-dissociation and have CO molecules,
(2) $X_{\rm CO}$ can be applied to a part (clump) of a cloud, and
(3) the gas is thermalized (the CO 3-2/1-0 brightness temperature ratio within the clump,
$R_{3-2/1-0}$, is unity).
In this case,
we should adopt $X_{\rm CO}=3\times 10^{20}$ from the Galactic measurement
even for the low metallicity,
because no part of the clump is photo-dissociated.
It gives
\begin{eqnarray}
    M_{\rm clump}
    &=& 3.6 \times 10^4 \Msun \left(\frac{S_{\rm CO(3-2)} dv}{1\,\rm Jy\cdot \kmps}\right)\left(\frac{D}{4.5\rm \, Mpc}\right)^{2}  \nonumber \\
    &\times & \left( \frac{R_{3-2/1-0}}{1.0} \right)^{-1} \left(\frac{X_{\rm CO(1-0)}}{3.0\times 10^{20}} \right). \label{eq:mclump3-2}
\end{eqnarray}
The $1\sigma$ sensitivity in the $2.54\kmps$ cube of CO(3-2) is $88\Msun$.
$M_{\rm clump}$ ranges from $M_{\rm clump}=2.7\times 10^2\Msun$ to $7.5\times 10^3\Msun$ (Table \ref{tab:clouds}).
Given the assumptions,
we expect the total uncertainties to be not smaller than a factor of 2-3.
Again, the clump mass in Cloud 4 is about the same as that of the OMC-1 clump
in the Orion A cloud.

\subsubsection{Consistency with CO(2-1) and Dust Measurements} \label{sec:mass21}

The detections in CO(3-2) are consistent with the previous non-detections in CO(2-1)
\citep{Bicalho:2019aa}.
The CO(3-2)-emitting clumps should also radiate the CO(2-1) emission as well,
but their CO(2-1) intensities are below the detection limit.
For example, Cloud 4, the brightest, has a total CO(3-2) flux
of $S_{\rm CO(3-2)} dv=207.7\,\rm mJy \kmps$
over the velocity width of $12.7\kmps$ (5 channels).
Adopting the Rayleigh-Jeans approximation for thermalized gas,
it translates to a total CO(2-1) flux of $S_{\rm CO(2-1)} dv=92.3\,\rm mJy \kmps$.
From Table \ref{tab:data},
the $1\sigma$ sensitivity in CO(2-1) over the same velocity width is $60.2\,\rm mJy \kmps$.
The brightest cloud/clump is only at a $1.5\sigma$ significance in the CO(2-1) measurement,
explaining the non-detection in CO(2-1).

If Cloud 4, the most massive,
were to be filled with CO as clouds in high metallicity (Figure \ref{fig:schematic}a),
its mass of $2.3\times 10^4\Msun$ would emit a total CO(2-1) flux of
$S_{\rm CO(2-1)} dv=200\,\rm mJy \kmps$,
using $X_{\rm CO(1-0)}=3.0\times 10^{20}$
in the high-metallicity regime and a CO(2-1) to CO(1-0) brightness ratio of
$R_{2-1/1-0}\sim 0.7$ for the bulk gas \citep{Braine:1992lr, Hasegawa:1997lr, Sorai:2001lr, Koda:2012lr, Koda:2020wt}.
This should have been detected at $3.3\sigma$ in the previous CO(2-1) measurement.

Note that the two assumptions we adopted above -- CO being confined only in the clump and being thermalized --
\textit{minimize} the predicted CO(2-1) flux for the detected CO(3-2) flux.
If the CO(2-1) emission is also from a cloud envelope,
it should add to the CO(2-1) flux.
If the gas is not thermalized, the CO 3-2/2-1 flux ratio would be lower,
and the predicted CO(2-1) flux would be higher.
Either way, the CO(2-1) emission from Cloud 4 should have been detected.
In other words, it is difficult to deviate significantly from the conditions we assumed.

The CO(2-1) non-detection also excludes the possibility of a CO-bright envelope
as large as Orion A's ($\sim20\pc$).
The CO(2-1) data can resolve such an envelope at the beam size of $\sim 17\times12\pc^2$ 
at a high sensitivity of 58~mK ($1\sigma$)  (Table \ref{tab:data}).
The non-detection suggests that its average surface brightness is $<0.17$~K ($3\sigma$).
Hence, the envelope is CO(2-1)-dark.

An independent constraint on $M_{\rm cloud}$ may be derived from the dust continuum emission.
Using the calibration by \citet{Scoville:2016aa} in a range of environments
and adopting the power-law dependence of the dust emissivity on frequency $\nu$ as $\kappa_{\nu} \propto \nu^{\beta}$,
we obtain
\begin{eqnarray}
    M_{\rm cloud}
    &=& 4.8 \times 10^8 \Msun \left( \frac{S_{\nu}}{\rm Jy} \right) \left(\frac{D}{4.5\rm \, Mpc}\right)^{2} \left( \frac{X_{\rm CO(1-0)}}{3.0\times 10^{20}} \right)  \nonumber \\
    &\times& \left( \frac{\nu}{\nu_{850\mu m}} \right)^{-(\beta+2)}
    \left( \frac{T_{\rm d}}{25\,\rm K} \right)^{-1}
    \left( \frac{\Gamma_{\nu}(T_{\rm d})}{\Gamma_{\nu850\mu m}(25{\rm K})}\right)^{-1}, \label{eq:mcloudcont}
\end{eqnarray}
where
\begin{equation}
    \Gamma_{\nu}(T_{\rm d}) = \frac{x}{e^x -1}, \text{ where } x=\frac{h\nu}{k_{\rm B}T_{\rm d}},
\end{equation}
and $\nu_{850\mu m}\approx 353$~GHz.
The $X_{\rm CO}$ term is due to the calibration in high metallicity.
We adopt a dust emissivity index of $\beta=1.8$ \citep{Planck-Collaboration:2011qy}
and dust temperature of $T_{\rm d}=25\,\rm K$ \citep{Scoville:2016aa}.
The detection limits from the 225 and 349~GHz dust continuum
measurements correspond to
$1.5\times 10^6\Msun$ and $4.5\times 10^4\Msun$ ($3\sigma$),
respectively.
The $M_{\rm cloud}$ from CO(3-2) is consistent with these limits (Table \ref{tab:clouds}).

\subsection{Star Formation On 1-kpc Scale}\label{sec:sf1kp}

On a $1\kpc$-scale average,
the SF efficiency in total gas content (HI+H$_2$) is much lower in XUV disks than
in the optical disks of nearby galaxies.
The average gas consumption timescale by SF,
i.e., the inverse of SF efficiency, is about 2~Gyr in the optical disks
(\citealp{Bigiel:2008aa, Leroy:2008fj}; see also \citealp{Kennicutt:1998kk, Kennicutt:2012aa}),
while it is as long as 100~Gyr over the XUV disk of M83 \citep{Bigiel:2010yq}.
As another reference, the gas consumption timescale in the Orion A cloud is about 520~Myr
(see Section \ref{sec:oriona} including the caveat on this estimation).

The region of this study is $1\kpc^2$ (more accurately, $0.99\kpc^2$ by taking into account the round corners of the observed region).
The total HI mass in the region is $M_{\rm HI}\sim 7.4\times 10^6\Msun$ from the VLA+GBT data.
The total H$_2$ mass is derived by summing up all the clouds
(Table \ref{tab:clouds}) and is $M_{\rm H_2}\sim 8.4\times 10^4\Msun$,
which includes the CO-dark H$_2$ gas.
Only $\sim 1\%$ of the gas is in the molecular phase.
The stellar initial mass function (IMF) may not be fully populated in this
low SF region \citep{Koda:2012ab}.
Nevertheless, if we simplistically adopt the standard calibration
that assumes a fully populated IMF \citep{Kennicutt:2012aa},
$L_{\rm H\alpha}\sim 6.8\times 10^{37}\,\rm erg\,s^{-1}$ (Section \ref{sec:Ha})
translates to a SFR of $\sim 3.7\times 10^{-4} \,\Msun\,{\rm yr}^{-1}$.

The atomic and molecular gas in total, including Helium (applying a factor of 1.36 to the mass),
would be consumed in about 27 Gyr at this SFR.
This is a few times shorter than the average across M83's XUV disk,
but is an order of magnitude longer than the average of 2~Gyr in the optical disks
and is longer than the Hubble time.
On the other hand, 
the consumption timescale of the molecular gas alone would be as short as 310~Myr,
which is similar to that in Orion A (Section \ref{sec:oriona}).

Therefore, the molecular clouds in the XUV disk on average appear to
be forming stars as much as
a typical cloud in the MW, but the XUV disk as a whole is not producing stars as much.
This supports the hypothesis that the clouds in the XUV disk and Galactic disk share
a similar mass structure, and hence, SF activity.
The key to the low SF activity in XUV disks should be in the conversion
efficiency from the atomic gas to molecular clouds.
This is a possibility suggested by \citet{Bigiel:2010yq},
and the resolved analysis of the clouds in this study supports it.

\subsection{Comparisons to Previous Studies} \label{sec:prev}

CO emission was previously detected in the outskirts of four galaxies with XUV disks.
Three of them, NGC 4414, NGC 6946, and M63, were observed with a single-dish telescope alone
at a relatively low sensitivity and resolution
\citep[$\gtrsim 10^{5-6}\Msun$ and $\gtrsim$300-500~pc, ][]{Braine:2004aa, Braine:2007aa, Dessauges-Zavadsky:2014fk}.
If our $1\kpc^2$ region in M83 were to be observed at these resolutions,
multiple clouds would be in a single telescope beam (Figure \ref{fig:maps}).
A detection in M33 \citep{Braine:2010aa}
was resolved into a single cloud of $\sim 4\times 10^4\Msun$ with an interferometer \citep{Braine:2012aa}.
Our study, together with \citet{Braine:2012aa}, demonstrates that
high resolution and sensitivity are necessary to identify individual clouds and
to investigate the environment of SF in the XUV disks.

There are large-scale CO(1-0) surveys in the outer disk of the MW
 \citep[e.g., ][]{Wouterloot:1989wu, Digel:1994tv, Brand:1994vn, Snell:2002tl, Yang:2002ua, Brunt:2003vd, Nakagawa:2005tk, Sun:2015wr}.
The optical edge of the MW disk is around $R_{25}\sim 13.4\kpc$ \citep{Goodwin:1997vl, Goodwin:1998tc}
with considerable uncertainty due to its edge-on projection.
The galactocentric radius of $1.24R_{\rm 25}$, corresponding to the $1\kpc^2$ region in M83,
is $r_{\rm gal}\sim 16.7\kpc$ for the MW.
The largest, most recent CO survey by \citet{Sun:2015wr} detected clouds in the far outer MW disk ($r_{\rm gal}\gtrsim 14\kpc$),
which include 51 clouds of $10^{2-4}\Msun$ in the extreme outer disk ($r_{\rm gal}\gtrsim 16$ up to $21\kpc$)
with 11 of them in the most massive end of $\gtrsim 10^4\Msun$.
They are spread over a large area of $\sim 80\kpc^2$ in the Galactic longitude of $l\sim$100-150$\arcdeg$.
Therefore, the cloud density is low.
The majority are associated with a possible extension of the Scutum-Centaurus
spiral arm traced in HI \citep{Sun:2015wr}.
SF exists even in the extreme outer disk,
but they appear to be mostly low-mass SF (later than B0-type)
according to an analysis of the Wide-field Infrared Survey Explorer (WISE) data \citep[][ see their Figure 15]{Izumi:2017wv}, while much obscured higher-mass SF has yet to be found.

Our $1\kpc^2$ region around an HI peak in the XUV disk of M83 has
23 clouds of $10^{3-4}\Msun$, including three $\gtrsim 10^4\Msun$ clouds.
Compared to the extreme outer MW disk, the cloud density is high.
Obviously, a much larger sample is required to make a general conclusion,
but the new detection appears to indicate that molecular clouds,
especially their abundance, are one of the key factors for the XUV activity.
Figure \ref{fig:scaling} shows clouds in the inner, outer, and far outer MW disk 
traced in CO(1-0) \citep{Solomon:1987pr, Sun:2015wr, Sun:2017vv}.
The XUV clouds in M83, traced in CO(3-2), generally show larger velocity dispersions
than the MW clouds.
This can be explained if the CO(3-2) emission primarily traces the dense clumps embedded in the clouds (Section \ref{sec:veldisp}).

\section{Implications}\label{sec:implication}

\subsection{Clouds and Star Formation in XUV Disks} \label{sec:cloudsandSF}

The ALMA CO(3-2) observations constrain the properties of 23 molecular clouds in the XUV disk,
having relatively small bulk gas masses of $M_{\rm cloud}=8.2\times 10^2$ to $2.3\times 10^4\Msun$,
and dense clump masses of $M_{\rm clump}=2.7\times 10^2$ to $7.5\times 10^3\Msun$.
The ubiquity suggests that these relatively low-mass clouds are a common population of molecular clouds
and are the main drivers of the SF activity in the XUV disk.

Numerous previous CO(1-0) and CO(2-1) searches in XUV disks resulted in non detection \citep{Watson:2017aa}.
A small number of previous detections could not isolate individual clouds
due to a combination of low sensitivity and resolution ($\gtrsim 10^{5-6}\Msun$ and $\gtrsim300$-$500\pc$),
and one successful case, in the very close galaxy M33, identified only one single cloud
of $4.3\times 10^{4}\Msun$\citep{Braine:2012aa}.
The common cloud population turns out to be less massive than the previous detections.
This study is the first demonstration of the utility of CO(3-2) to detect molecular clouds.

A comparison to Galactic clouds provides insights into the clouds and SF in the XUV disk.
For example, a cloud like Orion A
-- having $M_{\rm cloud}=4.0\times 10^4\Msun$ and $M_{\rm clump}=7.4 \times 10^3\Msun$ --
can be one of the most massive clouds in the observed region.
The SF activity in the region, traced by H$\alpha$, is also similar to
that of the Orion Nebula in Orion A.
Orion A is one of the least impressive massive star-forming regions in the MW,
but would be the most impressive in the XUV disk.
Clouds smaller than Orion A are also found in the XUV disk.
 
It is likely, and we hypothesize, that the internal mass distributions
of molecular clouds are similar between the MW disk and XUV disks.
The Galactic clouds, including Orion A,
typically have star-forming dense clumps surrounded by thick layers of bulk molecular gas.
In the case of low metallicity of XUV disks, the outer layers should have CO molecules
photo-dissociated and become CO-deficient and CO-dark \citep{Maloney:1988lr, van-Dishoeck:1988br, Wolfire:2010aa}.
We further hypothesize that
the dense clumps at the hearts of the clouds remain intact and maintain a high CO abundance
even in the low metallicity (Figure \ref{fig:schematic}).
This structure consistently explains the detection of the CO(3-2) emission mainly from the dense clumps,
and previous non-detection of the CO(2-1) emission.
The common clump-envelope structure between the Galactic and XUV clouds is expected
if they are governed primarily by their internal physics (e.g., by self-gravity)
independent of their environment.
This hypothesis can be tested with future observations, e.g.,
higher-resolution CO(3-2) observations for resolving the compactness of the clumps,
deeper CO(2-1) or CO(1-0) observations for confirming the thermalization,
and deeper dust continuum observations for independent mass estimation.

The common mass structure of clouds suggests that average clouds of the same mass should form stars
at a similar rate independent of their environment.
In other words, the SF efficiency on cloud scales,
on average, would be universal.
This is supported by the similar gas consumption timescales
between the XUV clouds and Orion A.

This study finds only relatively low-mass molecular clouds in the XUV disk of M83.
The absence of very bright HII regions in XUV disks 
may be due to the lack of more massive clouds.
\citet{Koda:2012ab} took the standard stellar initial mass function \citep{Salpeter:1955uq}
as the probability distribution function of stellar mass,
and calculated that a star cluster should have a total stellar mass
of at least $10^3\Msun$ to populate one $40\Msun$ star \citep[O6 or O7-type;][]{Sternberg:2003yb}.
The most massive cloud in our sample has $M_{\rm cloud}=2.3 \times 10^4\Msun$,
requiring a cloud-to-star mass conversion efficiency of $\sim 4\%$
to produce a $10^3\Msun$ cluster.
This is about the efficiency in Galactic molecular clouds \citep{Enoch:2007pb}.
The other clouds have even smaller $M_{\rm cloud}$ and
are unlikely to form very massive stars and
prominent HII regions.

The low SF efficiency on a 1~kpc scale has to originate
on scales larger than the clouds;
for example, in the conversion efficiency of the extended HI gas
into molecular clouds \citep{Bigiel:2010yq}.
The mechanism for atomic to molecular gas conversion
should also explain the cloud mass function
and the lack of massive molecular clouds.
Further studies are required to elucidate these aspects,
but some clue may already be in the distribution of the CO(3-2) clouds/clumps.
The long, chain-like distributions over $600\pc$,
and potentially the velocity gradient across our field,
indicate that
their formation is triggered on large scales,
such as by large-scale galactic flows.

The hypothesis of the common cloud mass structure (i.e., the clump-envelope structure)
in conjunction with the selective photo-dissociation
explains the clouds and SF activity
over the optical and XUV disks.
This simplistic view could be modified in future studies.
For example, this hypothesis exhibits merely an average picture
over cloud populations and evolutionary sequences
\citep[e.g., see ][]{Vazquez-Semadeni:2018vw}.
In other extreme environments,
such as in the dense Galactic center region,
the cloud properties show environmental influences \citep[e.g., ][]{Oka:2001qt}
although such regions are relatively small in volume and mass.

\subsection{Application to High-z and Other Environments}

The common cloud structure (Figure \ref{fig:schematic}) suggests
a constant $M_{\rm cloud}/L_{\rm CO(3-2)}$
between high- and low-metallicity environments,
even when CO molecules in cloud outer layers are photo-dissociated.
This has an important implication for CO observations of distant galaxies.
The red-shifted high-$J$ transitions of CO (higher than $J$=1 or 2)
are often used to trace
their total gas mass \citep[][]{Carilli:2013vk, Combes:2018we}.
Those high-$J$ CO emissions likely capture only star-forming dense clumps,
but not the bulk cloud gas.
In fact, the CO(1-0) and CO(2-1) luminosities are
often beyond those expected from higher-$J$ transition analysis
\citep{Ivison:2011uw, Carilli:2011wk, Riechers:2011ti, Bothwell:2013wb, Daddi:2015wk}
Even then, the high-$J$ CO emissions can trace the total gas mass
if the self-gravitating clouds have the same internal structure between
the local and distant universe,
and if the average ratio between the dense clumps and their parent clouds is
about constant independent of environment and redshift
as expected for self-gravitating clouds.

\section{Summary}

The new ALMA CO(3-2) observations detected a large number of molecular clouds
across a $1~\kpc^2$ area of the XUV disk of M83.
Their high abundance suggests that the 23 clouds represent
a common population of molecular clouds
and are likely the main drivers of the SF activity in XUV disks.

We hypothesized that these clouds share, on average, the same common structure (mass distribution) as Galactic clouds, such as Orion A;
having star-forming dense clumps embedded in thick layers of bulk molecular gas.
In the low-metallicity regime of XUV disks, the outer layers of the clouds are likely
CO-deficient and CO-dark, while still abundant in H$_2$,
due to selective photo-dissociation.
This hypothesis consistently explains the CO(3-2) detection and CO(2-1) non-detection.
The CO(1-0) and CO(2-1) emission should be reduced significantly
due to the lack of CO molecules in the photo-dissociated outer layers
where a large fraction of the CO(1-0) and CO(2-1) emission normally originate.
The dense clumps have sufficiently high density and temperature for
CO(3-2) excitation, reside at the hearts of the clouds, and thus, are
protected from the dissociation.
The common cloud mass structure independent of environment
is conceivable if it is governed by their internal physics, e.g., by self-gravity,
while their chemical structure could still depend on the environment (e.g., metallicity and radiation field).

The common cloud structure
constrains the geometry and physical conditions within molecular clouds,
which permits us to estimate their masses.
The total masses of the clouds, including the CO-dark H$_2$ component,
ranges from $M_{\rm cloud}=8.2\times 10^2$ to $2.3\times 10^4\Msun$.
Their star-forming dense clumps have masses of
$M_{\rm clump}=2.7\times 10^2$ to $7.5\times 10^3\Msun$.
The most massive clouds and their dense clumps are
similar to Orion A and its dense clump OMC-1.
The dense clumps should also radiate CO(2-1) emission, but
the flux is lower than the detection limit of the previous CO(2-1) study.
This study demonstrates, for the first time, the utility of CO(3-2) emission to find molecular clouds.

The SF activity and efficiency are also similar
between the clouds in the XUV disk and Orion A.
The XUV disk as a whole shows a much lower SF activity,
which is not due to peculiar cloud properties,
but due to the lack of massive clouds.
The abundance and mass function of molecular clouds
should be regulated by the (unknown) trigger of cloud formation from the abundant atomic gas.
Further studies are necessary to elucidate this mechanism,
but a clue is already found in the distribution of the clouds.
The long chain-like distribution ($\gtrsim 600\pc$ in length)
indicates that the trigger is on large scales.

The common mass structure and SF efficiency among molecular clouds 
should justify the use of high-$J$ CO transitions
as tracers of bulk molecular gas mass in distant galaxies,
assuming that the bulk molecular gas mass is proportional to
the dense clump mass traced by the high-$J$ transitions.

\begin{acknowledgments}

We thank the referees for useful comments, in particular, the first referee for suggesting the discussion in Section \ref{sec:prev} and the second referee for the advice on the title.
This paper makes use of the following ALMA data: ADS/JAO.ALMA\#2013.1.00861.S and 2017.1.00065.S.
ALMA is a partnership of ESO (representing its member states), NSF (USA) and NINS (Japan), together with NRC (Canada), MOST and ASIAA (Taiwan), and KASI (Republic of Korea), in cooperation with the Republic of Chile. The Joint ALMA Observatory is operated by ESO, AUI/NRAO and NAOJ.
The Green Bank Observatory and the National Radio Astronomy Observatory are facilities of the National Science Foundation operated under cooperative agreement by Associated Universities, Inc..
This research is based in part on data collected at the Subaru Telescope, which is operated by the National Astronomical Observatory of Japan (NAOJ). We are honored and grateful for the opportunity of observing the Universe from Maunakea, which has the cultural, historical, and natural significance in Hawaii.
Data analysis is in part carried out on the Multi-wavelength Data Analysis System operated by the Astronomy Data Center (ADC) in NAOJ.
JK acknowledges support from NSF through grants AST-1812847 and AST-2006600 and from NASA grant NNX14AF74G.
MR wishes to acknowledge support from ANID(CHILE) through FONDECYT grant No1190684.

\end{acknowledgments}

\facilities{ALMA, Subaru, GBT, VLA, GALEX}
\software{CASA 5.8, GBTIDL, Kapteyn Package, Clumpfind}

\end{document}